\documentclass[10pt,journal,twocolumn]{IEEEtran}

\usepackage{cite}

\ifCLASSINFOpdf
  \usepackage[pdftex]{graphicx}
\else
  \usepackage[dvips]{graphicx}
\fi

\usepackage[cmex10]{amsmath}

\usepackage{algorithmic,algorithm}
\usepackage[tight,footnotesize]{subfigure}
\usepackage{DefineNotations}
\usepackage{amsfonts}
\usepackage{amsbsy}
\usepackage{color}

\begin{document}
\title{WSR Maximized Resource Allocation in Multiple DF Relays Aided OFDMA Downlink Transmission}
\author{Tao Wang and Luc Vandendorpe,~\IEEEmembership{Fellow, IEEE}
\thanks{
Copyright (c) 2011 IEEE. Personal use of this material is permitted.
However, permission to use this material for any other purposes must
be obtained from the IEEE by sending a request to pubs-permissions@ieee.org.

T. Wang and L. Vandendorpe are with
Communications and Remote Sensing Laboratory,
Institute of Information and Communication Technologies, Electronics and Applied Mathematics (ICTEAM), Universit\'e Catholique de Louvain,
Louvain-la-Neuve, Belgium
(email: \{tao.wang,luc.vandendorpe\}@uclouvain.be).
}
\thanks{
The authors would like to thank the Walloon Region for funding the projects
MIMOCOM, the ARC SCOOP and the FP7 Network NEWCOM++.
}
}

\markboth{DRAFT PAPER}%
{Shell \MakeLowercase{\textit{et al.}}: Bare Demo of IEEEtran.cls for Journals}



\maketitle

\begin{abstract}
This paper considers the weighted sum rate (WSR) maximized resource allocation (RA)
constrained by a system sum power in an orthogonal frequency division multiple access (OFDMA)
downlink transmission system assisted by multiple decode-and-forward (DF) relays.
In particular, multiple relays may cooperate with the source
for every relay-aided transmission.
A two-step algorithm is proposed to find the globally optimum RA.
In the first step,
the optimum source/relay power and assisting relays that maximize the rate is found
for every combination of subcarrier and destination,
assuming a sum power is allocated to the transmission at that subcarrier
to that destination in the relay-aided transmission mode and the direct mode, respectively.
In the second step,
a convex-optimization based algorithm is designed to find the globally optimum assignment
of destination, transmission mode, and sum power for each subcarrier to maximize the WSR.
Combining the RAs found in the two steps, the globally optimum RA can be found.
In addition, we show that the optimum RA in the second step can readily be derived
when the system sum power is very high.
The effectiveness of the proposed algorithm is illustrated by numerical experiments.
\end{abstract}


% \fi
\IEEEpeerreviewmaketitle

\section{Introduction}

{Relay-aided cooperative wireless transmission has been attracting intensive
research interest lately, motivated by the consideration that by cooperative relaying,
distributed multiple input and multiple output (MIMO) links can be created for performance improvement,
which is suitable especially for applications
incapable of installing multiple antennas at the same radio device \cite{Pabst04}.
Two protocols, namely amplify and forward (AF) as well as decode and forward (DF),
have been proposed and become the focus of recent research works \cite{Laneman04}.
{In this paper, we consider half-duplex radio devices adopting the DF protocol,
which carries out a relay-aided transmission in two time slots,
referred to as the broadcasting slot and the relaying slot, respectively.}
In the broadcasting slot, the source emits a symbol, which is received
by both the destination and the relays.
In the relaying slot, some relays first recover the source symbol from received signals,
then transmit that recovered symbol to the destination.
Finally, the destination combines the signals received in the two slots,
then decode for the source symbol.
In particular, it has been show in \cite{Laneman04} that selection relaying DF,
which uses either the direct transmission mode without any relay assisting,
or the relay-aided mode depending on channel state information (CSI), achieves full diversity.
}

This paper addresses the resource allocation (RA) for an orthogonal frequency division multiple access (OFDMA)
downlink transmission system aided by multiple relays adopting selection relaying DF.
The motivation behind considering OFDMA is that
it is a widely recognized multiuser transmission technique for current and future wireless systems,
thanks to its flexibility to incorporate dynamic RA for performance improvement \cite{Wang11}.
Compared to the conventional OFDMA transmission, the relay-aided OFDMA transmission
raises more complicated RA problems, since it introduces extra tasks
such as deciding the transmission mode of each subcarrier,
determining assisting relays and the power allocation to them for each relay-aided subcarrier,
besides assigning destination and source power at each subcarrier.
Therefore, novel efficient RA algorithms are solicited for relay-aided OFDMA transmission systems.

For the point to point OFDM transmission aided by DF relays,
some RA algorithms have been proposed lately.
To name a few, RA algorithms have been proposed in
\cite{Gui08} to minimize the sum power under rate constraints,
and in \cite{Li08,Vandendorpe08-1} to maximize the sum rate subject to power constraints,
when only one relay exists and assists with selection relaying DF.
However, at every subcarrier assigned to the direct transmission mode,
the source does not transmit any symbol in the relaying slot, which wastes bandwidth resource.
To address this issue, rate-optimized RA algorithms which
permit source transmission in the relaying slot at each subcarrier assigned to the direct mode,
have been proposed in \cite{Vandendorpe08-2,Vandendorpe09-1,Vandendorpe09-2,Vandendorpe09-3,Vandendorpe09-4,WangJSAC11}.

For the OFDMA downlink transmission aided by DF relays,
some RA algorithms have been proposed as well.
For example, power constrained RA algorithms proposed in
\cite{Nam07,Kaneko07,Kwak07} and \cite{Cui09}
consider respectively the maximization of the sum rate and the weighted sum goodput,
whereas the one in \cite{Salem10} aims at maximizing a metric
depending on the rates and queue lengths of the source and relays.
Using those algorithms, each destination may decode the source transmitted signals,
each at a distinct subcarrier in the broadcasting and relaying (except for \cite{Cui09}) slots,
as well as the relays transmitted signals,
each at a distinct subcarrier from a single relay in the relaying slot.
Note that when the source transmits signals to relays in the broadcasting slot,
every destination discards the received signal replicas and thus spatial diversity is not exploited.
In \cite{Ng07}, a RA algorithm is proposed to maximize
the sum utility of multiple uplink/downlink data streams
aided by a single destination adopting selection relaying AF or DF.

\begin{figure}[!t]
  \centering
  \subfigure[]{
     \includegraphics[width=1in, height = 1.3in]{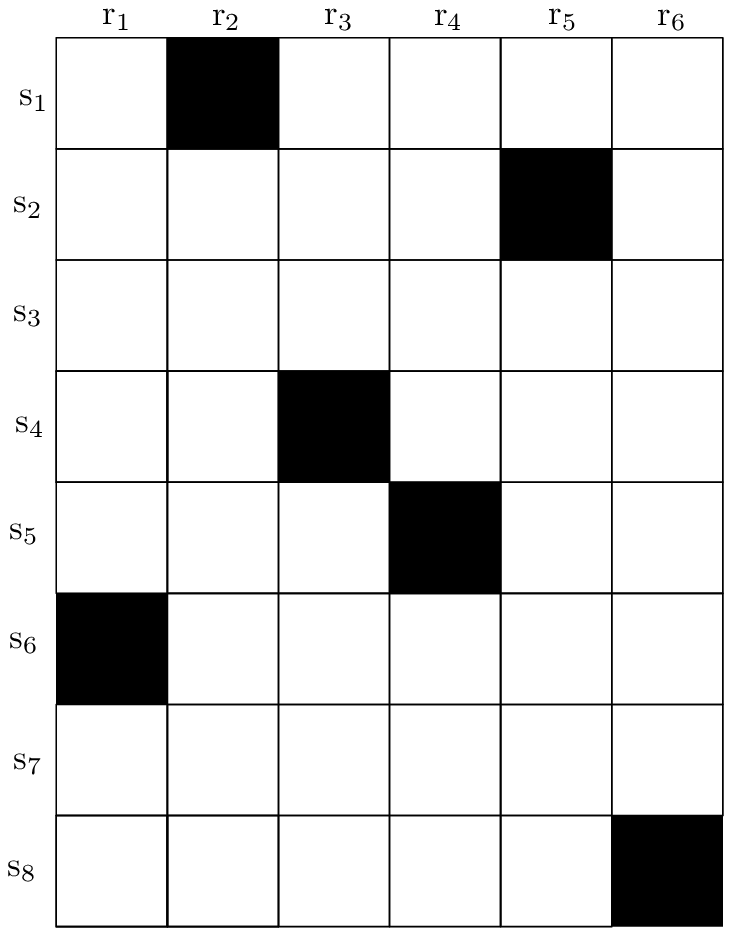}}
  \hspace{0.3in}
  \subfigure[]{
     \includegraphics[width=1in, height = 1.3in]{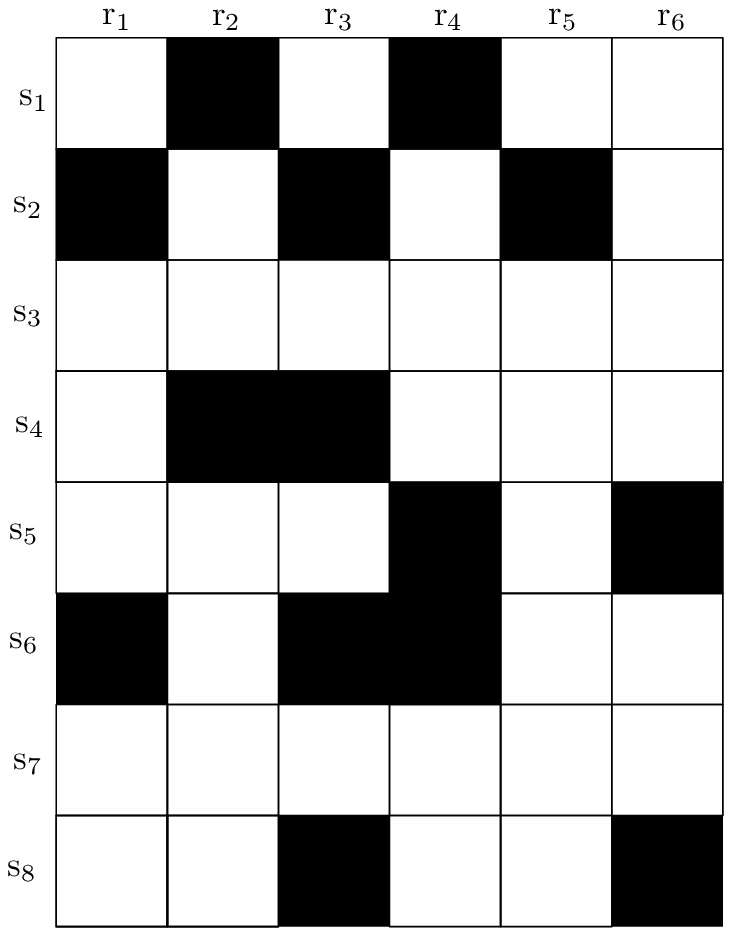}}
  \caption{Example patterns of selecting
           (a) one relay or (b) multiple relays assisting
           at each relay-aided subcarrier,
           where a black block at the $i$-th row and $j$-th column indicates
           the $j$-th relay ${\rm r}_j$ assists the transmission at the $i$-th subcarrier ${\rm s}_i$.
           }
  \label{fig:RelayPattern}
\end{figure}

So far, the majority of the proposed RA algorithms, as the aforementioned ones,
restrict that at most one relay can assist the source for every relay-aided transmission.
In fact, when there are multiple relays available,
allowing not just one but {\it each} of them to be eligible for assisting at every subcarrier,
can better exploit the degrees of freedom in the system for performance improvement.
For illustration purposes, example patterns of selecting single or multiple
assisting relays are shown in Figure \ref{fig:RelayPattern}.
In fact, these benefits have been demonstrated in \cite{Vandendorpe09-3}
for the OFDM transmission to a single destination, and in \cite{Kadloor10,Vardhe10}
for the OFDMA downlink transmission with each destination using only one fixed subcarrier.

Compared to the above existing works, this paper contains the following contributions:
\begin{itemize}
\item
we consider the weighted sum rate (WSR) maximized RA problem
constrained by a system sum power in an OFDMA downlink transmission system aided by multiple DF relays.
In particular, multiple relays may cooperate with the source for every relay-aided transmission.
\item
we propose a two-step algorithm to find the globally optimum RA based on a divide-and-conquer strategy.
In particular, we show that the optimum RA in the second step can be easily derived
when the system sum power is very high.
\end{itemize}

The remainder of this paper is organized as follows.
In Section II, we describe the considered system.
We propose the two-step RA algorithm in Section III,
then derive the optimum RA for the second step when the system sum power is very high in Section IV.
In Section V, the effectiveness of the proposed algorithm is illustrated by numerical experiments.
Finally, some conclusions complete this paper in Section VI.

\section{System description and RA problem}

We consider an OFDMA downlink transmission system from a source to
$U$ destinations aided by $N$ DF relays collected
in the set $\Rset=\{\Relayi| i=1,\cdots,N\}$.
All links are assumed to be frequency selective, and OFDM with sufficiently long
cyclic prefix is used to transform every link into $K$
parallel channels, each at a different subcarrier facing flat fading.
At each subcarrier, the transmission of a symbol is in either the direct mode,
or the relay-aided mode spanning across two equal-duration time slots,
namely the broadcasting slot and the relaying slot.
Due to the OFDMA, each subcarrier is allocated to one destination exclusively.
{As will be described later, in the broadcasting slot only the source transmits symbols at all subcarriers.
In the relaying slot, the source transmits at the subcarriers assigned to the direct mode,
whereas some relays transmit simultaneously at the other subcarriers assigned to the relay-aided mode.

We make the following assumptions for the considered system.
First, we assume the radio frequency (RF) nonlinearity at the source and the relays is negligible,
and the carrier frequency and symbol timing of the source
are perfectly synchronized with those of the relays,
e.g. with the techniques in \cite{OFDMAsyn07}. }
Second, every channel in the system remains unchanged within a sufficiently long duration,
over which the RA algorithm can be implemented at a central controller
knowing precisely the CSI of the system.
Furthermore, the RA information can be reliably disseminated to the source, every relay, and every destination.
{To date, the related works, as introduced in Section I, are based on these assumptions as well.
As an interesting topic, the RA for the system with subcarrier nonorthogonality and non-ideal CSI
will be studied in our future work.}

% between any two of source, $\Relayi$, and $u$
\begin{table}[!b]
  \centering
  \caption{Channel coefficients at subcarrier $k$.}\label{tab:chcoeff}
  \begin{tabular}{|c|c|c|c|}
     \hline
       source to destination $u$   &  source to $\Relayi$  &  $\Relayi$ to destination $u$ \\
     \hline
       $\Cstuk$ & $\Cstik$ & $\Cituk$ \\
     \hline
  \end{tabular}
\end{table}

\begin{figure}[!t]
  \centering
  \subfigure[In the broadcasting slot]{
     \includegraphics[width=1.5in]{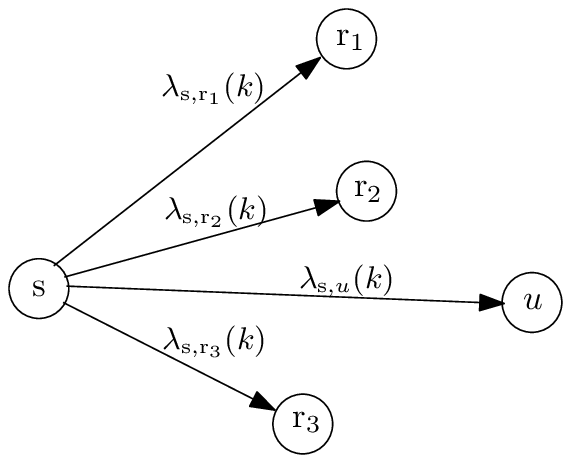}}
  \hspace{0.1in}
  \subfigure[In the relaying slot]{
     \includegraphics[width=1.5in]{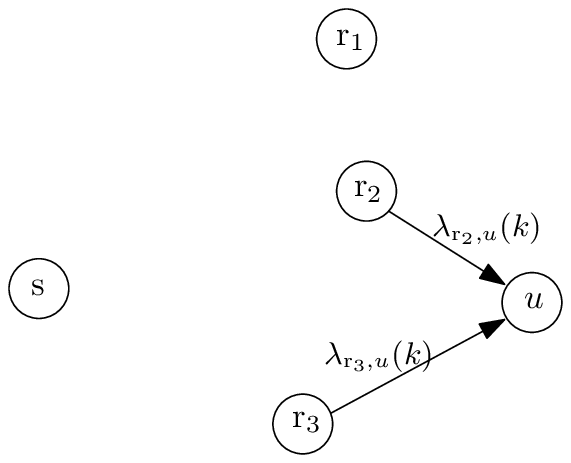}}
  \caption{Illustration of a relay-aided transmission at subcarrier $k$, where $\Rsetuk=\{\mathrm{r}_2,\mathrm{r}_3\}$.}
  \label{fig:RelayTx}
\end{figure}

Let's consider the transmission of a unit-variance symbol $s$ at a subcarrier to a destination,
say at subcarrier $k$ to destination $u$.
The coefficient of the channel between any two of the source,
$\Relayi$, and destination $u$, are notated according to Table \ref{tab:chcoeff}.
We first describe the transmission in the relay-aided mode.
Suppose a set of relays, collected in the set $\Rsetuk$, are selected by the RA algorithm to assist relaying.
Using the transmit power $\Psuk$,
the source first emits the symbol $\sqrt{\Psuk}{s}$, while each relay does not transmit anything
in the broadcasting slot, as illustrated in Figure \ref{fig:RelayTx}.a.
At the end of this slot,
both destination $u$ and every relay receive the source signal.
The signal samples at destination $u$ and $\Relayi$ can be expressed by
\begin{align}
    \YukB = \sqrt{\Psuk}\Cstuk{s} + \NukB
\end{align}
and
\begin{align}
    \YikB = \sqrt{\Psuk}\Cstik{s} + \NikB,
\end{align}
respectively, where $\NukB$ and $\NikB$ represent the corruption of
the additive white Gaussian noise (AWGN) at destination $u$ and $\Relayi$, respectively.

{After decoding and then reencoding $s$, all relays in $\Rsetuk$ transmit simultaneously
to destination $u$ in the relaying slot, which in effect establishes a distributed
multiple input and single output (MISO) transmission link as illustrated in Figure \ref{fig:RelayTx}.b.
Specifically, $\Relayi\in\Rsetuk$ transmits $\alpha_{\Relayi}s$,
where $\alpha_{\Relayi}$ is the complex weight for transmit beamforming
and satisfies $|\alpha_{\Relayi}|^2=\Piuk$,
with $\Piuk$ denoting the transmit power allocated to $\Relayi$.
To have the relays' signals add coherently when received at destination $u$,
$\alpha_{\Relayi} = \sqrt{\Piuk}e^{-j\arg{(\Cituk)}}$ is used,
where $\arg{(\Cituk)}$ stands for the phase of $\Cituk$.
Note that the above described relay-aided transmission enables a flexible use
of all relays opportunistically through a general form of adaptive transmit beamforming,
in that $\Rsetuk$ and $\{\Piuk | \Relayi\in\Rsetuk\}$ are determined dynamically
by the RA algorithm based on the CSI, as will be developed later.}
At the end of the relaying slot, the signal sample at destination $u$ is denoted by
\begin{align}
    \YukR = \sum_{\Relayi\in\Rsetuk}\sqrt{\Piuk}|\Cituk|s + \NukR,
\end{align}
where $\NukR$ represents the corruption of AWGN at destination $u$.

Finally, destination $u$ combines $\YukB$ and $\YukR$ based on the maximum ratio combining (MRC),
and the output is denoted by
\begin{align}
    \Yumrc = &\sqrt{\Psuk}(\Cstuk)^*\YukB +     \nonumber \\
             &\big(\sum_{\Relayi\in\Rsetuk}\sqrt{\Piuk}|\Cituk|\big)^*\YukR.
\end{align}

We assume $\{\NikB | \Relayi\in\Rsetuk\}$, $\NukB$, and $\NukR$ are independent
zero-mean circular Gaussian random variables with the same variance $\NVar$.
By some mathematical arrangements, the signal to noise ratio (SNR) associated with decoding 
$s$ from $\YikB$ at $\Relayi\in\Rsetuk$ in the broadcasting slot is
\begin{align}
    \SNRik = \Psuk\Gstik,
\end{align}
and the SNR associated with decoding $s$ from $\Yumrc$ at destination $u$ is
\begin{align}
    \SNRumrck  = &\Psuk\Gstuk +                                                \nonumber \\
                 &\big(\sum_{\Relayi\in\Rsetuk}\sqrt{\Piuk\Gituk}\big)^2,
\end{align}
where $\Gstik=\frac{|\Cstik|^2}{\NVar}$, $\Gstuk=\frac{|\Cstuk|^2}{\NVar}$,
and $\Gituk=\frac{|\Cituk|^2}{\NVar}$
represent the noise power normalized channel gains from
the source to $\Relayi$, from the source to destination $u$, and from $\Relayi$ to destination $u$, respectively.

To guarantee reliable decoding at destination $u$ and every relay in $\Rsetuk$,
the maximum achievable rate at subcarrier $k$ is
\begin{align}
    \RukR =& \min\{\min_{\Relayi\in\Rsetuk}\{\ln(1 + \SNRik)\}, \ln(1 + \SNRumrck)\}  \nonumber\\
          =& \ln(1 + \SNRuk) \label{eq:RukR}
\end{align}
in nats/two-slots, where
\begin{align}
    \SNRuk     = \min\{\SNRumrck,\SNRminRsetuk\} \label{eq:SNRuk}
\end{align}
and
\begin{align}
    \SNRminRsetuk = \Psuk\cdot\min_{\Relayi\in\Rsetuk}\Gstik. \label{eq:SNRminRsetuk}
\end{align}

As for the direct transmission, we consider a protocol with a better use of system bandwidth
compared to requiring the source to transmit only in the broadcasting slot
considered in \cite{Li08,Vandendorpe08-1}.
Specifically, the source emits two independent symbols with transmit powers
$\PsBuk$ and $\PsRuk$ in the two slots, respectively,
and only destination $u$ decodes the transmitted symbols from the two received signal samples.
We assume the AWGN corruption associated with the two received samples
are independent zero-mean circular Gaussian distributed with variance $\NVar$.
By simple mathematical arrangements, the sum achievable rate to destination $u$ in nats/two-slots
can be evaluated as
\begin{align}
    \RukD = \ln(1 + \PsBuk\Gstuk) + \ln(1 + \PsRuk\Gstuk).
\end{align}

We consider the RA problem of finding the optimum assignment of destination, transmission mode,
and source power for each subcarrier, as well as the optimum assisting relays and their transmit powers
for every relay-aided subcarrier, to maximize the WSR of the destinations
when the considered system consumes a sum power no greater than $\Ptot$.
{Note that for the system under consideration the same subcarrier
is used by the source and the relays for transmitting a symbol in the relay-aided mode.
This constraint could however be relaxed, and optimized subcarrier
pairing could also be implemented. This would further increase
the degrees of freedom for optimization.
However, it would be more involved to gain insights on the RA algorithm design,
and that is why in the current work, the RA algorithm is designed
under this constraint of nonoptimized subcarrier pairing.
We will nevertheless use the insights gained here to guide future work,
which will also consider subcarrier pairing.
}

\section{The two-step RA algorithm}\label{sec:II-b}

We propose a two-step algorithm to solve the RA problem based on a divide-and-conquer strategy.
We will first give an overview, then develop this algorithm in the following subsections.

\subsection{An overview of the two-step algorithm}

In the first step, the optimum $\Rsetuk$, $\Psuk$, and $\{\Piuk | \forall\,\Relayi\in\Rsetuk\}$
that maximize $\RukR$, and the optimum $\PsBuk$ and $\PsRuk$ that maximize $\RukD$,
are found for every combination of subcarrier $k$ and destination $u$,
assuming that a sum power $P$ is used by the transmission at subcarrier $k$ to destination $u$
in the relay-aided mode and the direct mode, respectively.
Such maximized $\RukR$ and $\RukD$ are denoted by $\RukR(P)$ and $\RukD(P)$, respectively.
The associated algorithm will be developed in Section III.B.

To simplify the RA in the second step,
two user sets, namely $\UDk=\{u|\forall\,P\in[0,\Ptot],\RukD(P)\geq\RukR(P)\}$
and $\URk=\{u|\forall\,P\in[0,\Ptot],\RukR(P)\geq\RukD(P)\}$,
are found for every subcarrier $k$.
Once a destination $u\in\UDk$ (resp. $u\in\URk$) is allocated with subcarrier $k$,
the direct mode (resp. the relay-aided mode) should always be used to maximize the WSR,
since it results in a rate no smaller than the relay-aided mode (resp. the direct mode),
independently of the sum power allocated to this subcarrier.

To formulate the maximum WSR with the $\RukR(P)$ and $\RukD(P)$ derived in the first step,
we define for every destination $u$ in $\URk$ (resp. $\UDk$)
a binary variable $\IukR$ (resp. $\IukD$) and a nonnegative variable $\PukR$ (resp. $\PukD$),
where $\IukR=1$ (resp. $\IukD=1$) indicates that subcarrier $k$ in the relay-aided (resp. direct) mode
is allocated to destination $u$,
and $\PukR$ (resp. $\PukD$) represents the corresponding sum power allocated to subcarrier $k$.
For every destination $u\notin\UDk\cup\URk$, $\IukR$, $\IukD$, $\PukR$, and $\PukD$ are
defined with the same interpretations as explained above.
Note that $\IukR$ and $\PukR$ are defined for every destination $u\notin\UDk$,
and $\IukD$ and $\PukD$ are defined for every destination $u\notin\URk$.
Now, the maximum WSR can be expressed by
\begin{align}
    f  = &\sum_{k=1}^K \big(\sum_{u\notin\UDk}\wu\IukR\RukR(\PukR) +        \nonumber \\
         &  \qquad \qquad \sum_{u\notin\URk}\wu\IukD\RukD(\PukD)\big) \label{eq:WSR}
\end{align}
where $\wu>0$ satisfying $\sum_{u=1}^U\wu=1$
is the weight assigned by system designers to the rate of destination $u$.
In particular, increasing $\wu$ leads to a higher priority given to destination $u$.

In the second step, for every subcarrier $k$ the optimum $\{\IukR,\PukR | \forall\,u\notin\UDk\}$ and
$\{\IukD,\PukD | \forall\,u\notin\URk\}$ that maximize $f$ are found subject to the system sum power constraint
\begin{equation}
  \sum_{k=1}^K(\sum_{u\notin\UDk}\IukR\PukR + \sum_{u\notin\URk}\IukD\PukD)\leq\Ptot, \label{eq:Ptot}
\end{equation}
as well as the constraint
\begin{align}
  \sum_{u\notin\UDk}\IukR + \sum_{u\notin\URk}\IukD \leq 1,   \forall\;k,      \label{eq:OFDMA}
\end{align}
due to the OFDMA.
It is important to note that there exist default constraints that
$\forall\;k,u$, $\Rsetuk$ should be a subset of $\Rset$,
and every power (resp. indicator) variable should be non-negative (resp. binary).
{This problem, consisting of both binary and continuous optimization variables,
is not convex since the feasible set for the binary variables is not convex.}
In Section III.C, we will develop an algorithm to find the globally optimum solution.
Combining the RAs found in the two steps, the globally optimum RA
for maximizing the WSR can be found.

\subsection{RA algorithm in the first step}\label{sec:step1}

First, we consider the maximization of $\RukD$ under the constraint $\PsBuk+\PsRuk=P$.
According to the Jensen's inequality, the maximum $\RukD$ in this case can be easily found as
\begin{equation}
    \RukD(P) = 2\ln\left(1 + \Gstuk\frac{P}{2}\right),
\end{equation}
and the optimum RA is to allocate the sum power $P$ equally to $\PsBuk$ and $\PsRuk$.

Next, we consider the maximization of $\RukR$ with respect to
$\Rsetuk$, $\Psuk$, and $\{\Piuk | \forall\,\Relayi\in\Rsetuk\}$ constrained by
\begin{align}
\Psuk + \sum_{\Relayi\in\Rsetuk}\Piuk=P.
\end{align}

To facilitate derivation, we define $\Gxtouk{\Rsetuk}=\sum_{\Relayi\in\Rsetuk}\Gituk$,
which represents the sum of the channel gains from all assisting relays to destination $u$.
$\SRsetk$ is defined as the set incorporating all relays
sorted in the increasing order of $\Gstik$,
and the $i$-th relay in $\SRsetk$ is denoted by $\SRelayki{i}$.
In particular, $\Gstoxk{\SRelayki{N}} = \max_{\Relayi\in\Rset}\Gstik$.
We define $\SRsetkitoN{i}$ as the set containing all relays in
$\SRsetk$ with indices from $i$ up to $N$.
When $\Gstuk < \Gstoxk{\SRelayki{N}}$,
$\xuk$ is defined as the smallest $i$ satisfying $\Gstuk<\Gstoxk{\SRelayki{i}}$.

{By means of intuitive figure illustrations,
the procedure of maximizing $\RukR$ is derived and put in Appendix A for clarity.
In particular, we find that
\begin{align}\label{eq:RukDPukD}
    \RukR(P) = \ln(1 + \GstukR{P}),
\end{align}
where $\GstukR$ can be evaluated by one of the following formulas:
\begin{enumerate}
\item
when $\Gstuk \geq \Gstoxk{\SRelayki{N}}$, $\GstukR = \Gstoxk{\SRelayki{N}}$.
In this case, $\forall\;P\in[0,\Ptot]$, $\RukD(P)\geq\RukR(P)$ holds.
\item
when $\Gxtouk{\SRsetkitoN{\xuk}} \leq \Gstuk < \Gstoxk{\SRelayki{N}}$, $\GstukR = \Gstuk$.
In this case, $\forall\;P\in[0,\Ptot]$, $\RukD(P)\geq\RukR(P)$ holds.
\item
when $\Gstuk < \Gstoxk{\SRelayki{N}}$ and $\Gstuk < \Gxtouk{\SRsetkitoN{\xuk}}$,
{\small
\begin{align}\label{eq:Gu1k}
\GstukR = \max_{\xuk\leq{b}\leq\yuk}\frac{\Gstoxk{\SRelayki{b}}\Gxtouk{\SRsetkitoN{b}}}{\Gxtouk{\SRsetkitoN{b}} + \Gstoxk{\SRelayki{b}} - \Gstuk},
\end{align}}
where $\yuk$ is the greatest $i$ satisfying $\Gxtouk{\SRsetkitoN{i}}>\Gstuk$.
In this case, $\GstukR>\Gstuk$ always holds as shown in Appendix A.
Suppose $b=\zuk$ is the maximizer for the right-hand side of \eqref{eq:Gu1k},
the optimum $\Rsetuk$ is $\SRsetkitoN{\zuk}$.
The optimum $\Psuk$ is computed by \eqref{eq:PsukR} with $\buk=\zuk$,
and the optimum $\Piuk$, $\forall\;\Relayi\in\SRsetkitoN{\zuk}$ is
\begin{align}\label{eq:PiukR}
\Piuk =\frac{(P-\Psuk)\Gituk}{\Gxtouk{\SRsetkitoN{\zuk}}}.
\end{align}
\end{enumerate}

The above analysis reveals that,
the relay-aided mode with the optimum RA in effect transforms all channels at subcarrier $k$
into a source to destination $u$ channel with normalized gain $\GstukR$ and half of the system bandwidth.
As a matter of fact, only one symbol can be sent during the two slots,
while the system bandwidth can actually support sending two independent symbols per two slots.
When $\GstukR\leq\Gstuk$, the direct mode should always be used.
When $\GstukR>\Gstuk$, using the relay-aided mode leads to an increased channel gain
but sacrificing half of the bandwidth, compared to using the direct mode.
In this case, $\RukR(P)\geq\RukD(P)$ holds if $P\leq\frac{4(G_{u,1}(k)-\Gstuk)}{(\Gstuk)^2}$, otherwise $\RukD(P)>\RukR(P)$,
which means that the relay-aided and direct modes
should be used for the low and high power regimes, respectively.
The interpretation is that,
in the low power regime, it is more beneficial
to increase the received power, while in the high power regime,
it is better to increase the number of channel uses per time unit \cite{Fund-WCOM}.
Based on the above analysis, $\UDk$ and $\URk$ for every subcarrier $k$
can be expressed as
\begin{align}
    \UDk = \{u|& \GstukR \leq \Gstuk\}     \label{eq:UDkcondition}
\end{align}
and
\begin{align}
   \URk = \{u|\GstukR>\Gstuk, \Ptot\leq\frac{4(G_{u,1}(k)-\Gstuk)}{(\Gstuk)^2}\}.   \label{eq:URkcondition}
\end{align}
}

\subsection{RA algorithm in the second step}\label{sec:step2}

We propose an algorithm to solve the RA problem in the second step
based on the following strategy.
First, for every subcarrier $k$, $\IukR$ and $\IukD$
are relaxed to be real variables between $0$ and $1$.
Now, $\IukR$ (resp. $\IukD$) can be interpreted as the fraction of
the whole duration allocated to transmitting at subcarrier $k$ to destination $u$
in the relay-aided (resp. direct) mode,
and $\EffRukR=\IukR\RukR(\PukR)$ (resp. $\EffRukD=\IukD\RukD(\PukD)$) can be interpreted as
the rate averaged over the whole duration for
transmitting to destination $u$ at subcarrier $k$ in the relay-aided (resp. direct) mode.
Then, the relaxed RA problem can be transformed with change of variables
into a convex optimization problem,
and its globally optimum RA can be found as will be shown later.
Most interestingly,
we will show that every optimum $\IukR$ and $\IukD$ for the relaxed problem are
still equal to either $0$ or $1$,
which means that the globally optimum RA to the relaxed problem
is also globally optimum to the original RA problem.

To make the change of variables, $\PukR$ is substituted with $\frac{\EukR}{\IukR}$ for every $u\notin\UDk$,
and $\PukD$ with $\frac{\EukD}{\IukD}$ for every $u\notin\URk$.
Now, the sum power constraint, $\EffRukR$, and $\EffRukD$ are respectively expressed as
\begin{align}
  \sum_{k=1}^K\big(\sum_{u\notin\UDk}\EukR + \sum_{u\notin\URk}\EukD\big)\leq\Ptot, \label{eq:Ptot-2}
\end{align}
\begin{align}
    \EffRukR &= \IukR\ln\left(1 + G_{u,1}(k)\frac{\EukR}{\IukR}\right), \label{eq:EffRukR}  \\
    \EffRukD &= 2\,\IukD\ln\left(1 + \frac{\Gstuk}{2}\frac{\EukD}{\IukD}\right) \label{eq:EffRukD}.
\end{align}

{Note that $\EffRukR$ (resp. $\EffRukD$) is a concave function
of $\IukR>0$ and $\EukR$ (resp. $\IukD>0$ and $\EukD$),
since it is a perspective of the concave function $\RukR(\PukR)$ (resp. $\RukD(\PukD)$)
(refer to page $89$ in \cite{Convex-opt} for more details).}
One delicate issue is that after the change of variables,
$\EffRukR$ (resp. $\EffRukD$) can not be evaluated at $\IukR=0$ (resp. $\IukD=0$).
To address this issue, we expand the domain of $\EffRukR$ to incorporate $\IukR=0$,
and define $\EffRukR=0$ when $\IukR=0$.
After the expansion and definition, $\EffRukR$ is still a concave and continuous
function of $\IukR\geq0$, because $\lim_{\IukR\rightarrow0}\EffRukR=0$.
Motivated by the same consideration,
we expand the domain of $\EffRukD$ to incorporate $\IukD=0$, and define $\EffRukD=0$ when $\IukD=0$,
so as to ensure $\EffRukD$ is a concave and continuous function of $\IukD\geq0$.

%---------------------------------------------------------------------------------------
%--------------------------------- dual-method-overview --------------------------------
%---------------------------------------------------------------------------------------

Based on the above analysis, the relaxed RA problem with new variables,
i.e., maximizing $f$ subject to the constraints \eqref{eq:OFDMA} and \eqref{eq:Ptot-2},
is a convex optimization problem.
Obviously, this problem satisfies the Slater constraints qualification,
i.e. there exists at least one feasible solution satisfying all inequality constraints strictly.
This justifies the zero duality gap for the problem \cite{Nonlinear-opt}.
Therefore, the dual method can be used to solve this problem for the globally optimum RA \cite{Yu06}.
Specifically, dual variables are defined for certain constraints,
then an iterative algorithm consisting of an inner loop and an outer loop is implemented.
In the inner loop,
the Lagrangian maximization problem (LMP), which is to find the optimum RA variables
that maximize the Lagrangian, is solved with the dual variables fixed.
In the outer loop,
the optimum dual variables are found by iteratively updating the dual variables with the subgradient method,
until the Karush-Kuhn-Tucker (KKT) conditions are satisfied.
The optimum RA variables for the LMP given the optimum dual variables,
are the globally optimum solution for the relaxed RA problem.

To use the dual method, a nonnegative dual variable $\mu$ is introduced for
the system sum power constraint \eqref{eq:Ptot-2}, and the associated LMP is
\begin{align}\label{eq:LagMax}
      \max\;  & L(\mu) = f + \mu\left(\Ptot - P_x\right)                  \nonumber\\
   {\rm s.t.} & \eqref{eq:OFDMA},                                                  \\
              & \IukR\geq0, \EukR\geq0, \forall\,k,\forall\,u\notin\UDk,  \nonumber\\
              & \IukD\geq0, \EukD\geq0, \forall\,k,\forall\,u\notin\URk,  \nonumber
\end{align}
where $L(\mu)$ represents the Lagrangian given $\mu$, and $P_x$ is the left-hand side of \eqref{eq:Ptot-2}.
{The optimum dual variable denoted by $\muo$, must satisfy the KKT conditions
consisting of (a) $\muo\geq0$, (b) $\muo(\Ptot-P_x(\muo))=0$,
and (c) $P_x(\muo)\leq\Ptot$,
where $P_x(\mu)$ represents the $P_x$ evaluated with the optimum power variables to \eqref{eq:LagMax} given $\mu$
\cite{Convex-opt,Nonlinear-opt}.
It is important to note that $\muo$ must be strictly positive, which can be justified
as follows.
Suppose $\muo = 0$. In this case, the maximum $L(\muo)$ subject to
the constraints of \eqref{eq:LagMax} is infinity,
since $f$, as a function of $\EffRukR$ and $\EffRukD$,
is monotonically increasing with $\EukR$ and $\EukD$, which can be increased unboundedly
while remaining feasible to \eqref{eq:LagMax}.
This means that $P_x(\muo)$ is infinity, which contradicts (c) of the KKT conditions.
Therefore, $\muo\neq0$, and the KKT conditions are reduced to $\mu>0$ and $\Ptot=P_x(\muo)$. }

%---------------------------------------------------------------------------------------
%-------------------------------------- inner-loop -------------------------------------
%---------------------------------------------------------------------------------------
Let's now find the optimum RA variables to \eqref{eq:LagMax} given $\mu>0$.
To this end, we find for every subcarrier $k$, $\forall\,u\notin\UDk$, the optimum $\EukR$ given $\IukR$,
and $\forall\,u\notin\URk$, the optimum $\EukD$ given $\IukD$, to maximize $L(\mu)$.
After mathematical arrangements, we can show these optimum values are
\begin{align}
   \rho_{u,k,i} = \left\{
                           \begin{array}{ll}
                               \IukR\left[\frac{\wu}{\mu} - \frac{1}{\GstukR}\right]^+    &  \mathrm{if}\,i=1,  \\
                              2\IukD\left[\frac{\wu}{\mu} - \frac{1}{\Gstuk}\right]^+     &  \mathrm{if}\,i=2,
                           \end{array}
                  \right.   \label{eq:optEukRD}
\end{align}
where $[x]^+ = \max\{x,0\}$.
Using \eqref{eq:optEukRD}, $L(\mu)$ is reduced to
\begin{align}
   L(\mu) = \sum_{k=1}^K L_k(\mu) + \mu\Ptot,
\end{align}
where
\begin{align}
   L_k(\mu) = \sum_{u\notin\UDk}{\IukR}X_{k}(u) + \sum_{u\notin\URk}{\IukD}Y_{k}(u),
\end{align}
and
\begin{align}
   X_{k}(u) = & \ln\left(1 +    \GstukR\left[\frac{\wu}{\mu}-\frac{1}{\GstukR}\right]^+\right) \nonumber\\
              & - \mu\left[\frac{\wu}{\mu}-\frac{1}{\GstukR}\right]^+,                    \label{eq:Xku}\\
   Y_{k}(u) = & 2\,\ln\left(1 + \Gstuk\left[\frac{\wu}{\mu}-\frac{1}{\Gstuk}\right]^+\right)   \nonumber \\
              & - \mu\left[\frac{2\wu}{\mu} - \frac{2}{\Gstuk}\right]^+.   \label{eq:Yku}
\end{align}

Now, the problem of finding the optimum $\{\IukR | \forall\,u\notin\UDk, \forall\,k\}$
and $\{\IukD | \forall\,u\notin\URk, \forall\,k\}$ for \eqref{eq:LagMax} can be decomposed
into $K$ subproblems, the $k$-th of which is to find the optimum
$\{\IukR | \forall\,u\notin\UDk\}$ and $\{\IukD | \forall\,u\notin\URk\}$
for maximizing $L_k(\mu)$ subject to the constraints in \eqref{eq:LagMax}.
It can be readily shown that
the optimum $\{\IukR | \forall\,u\notin\UDk\}\cup\{\IukD | \forall\,u\notin\URk\}$
has all entries equal to $0$, except for one entry $\Iukio$ equal to $1$.
Obviously, $(\uk,\ik)$ is the $(u,i)$ with the maximum metric $\metricuik$, expressed as
\begin{align} \label{eq:metricuik}
  \metricuik = \left\{
\begin{array}{ll}
X_{k}(u)    &       \mathrm{if}\,i=1,      \\
Y_{k}(u)   &        \mathrm{if}\,i=2.
\end{array}
\right.
\end{align}

Note that when computing $\metricuik$, $i$ can only be $1$ (resp. $2$)
if $u\in\URk$ (resp. $u\in\UDk$),
while $i$ may be either $1$ or $2$ if $u\notin\URk\cup\UDk$.
There might exist multiple combinations of $(u,i)$ corresponding to the maximum $\metricuik$.
In this case, any combination of them can be chosen to be $(\uk,\ik)$.
According to \eqref{eq:optEukRD}, the optimum power variables related to subcarrier $k$
are all equal to zero, except for $\Pukio$ computed with
\begin{align}
   \Pukio   = \left\{
                       \begin{array}{ll}
                          \left[\frac{w_{u_k}}{\mu} - \frac{1}{G_{u_k,1}(k)}\right]^+   &  \mathrm{if}\,\ik=1,  \\
                          2\left[\frac{w_{u_k}}{\mu} - \frac{1}{G_{s,u_k}(k)}\right]^+  &  \mathrm{if}\,\ik=2.
                       \end{array}
              \right.  \label{eq:optPukRD}
\end{align}

\begin{algorithm}
\caption{Procedure to solve \eqref{eq:LagMax}}\label{alg:LagMax}
\begin{algorithmic}
{\small
    \FOR{$k=1$ to $K$}
         \STATE $\forall\,u$ and the associated $i$, compute $\metricuik$ defined in \eqref{eq:metricuik};
         \STATE $\{\uk,\ik\}=\arg\max_{u,i}\metricuik$;
         \STATE $\Iukio=1$ and compute $\Pukio$ by \eqref{eq:optPukRD};
         \STATE $P_{u,k,i}=0$ and $t_{u,k,i}=0$ if either $u\neq\uk$ or $i\neq\ik$.
    \ENDFOR
}
\end{algorithmic}
\end{algorithm}

\begin{algorithm}
\caption{The RA algorithm in the second step}\label{alg:step-2}
\begin{algorithmic}
{\small
%\STATE $\forall\;u,\;k$, use the algorithm in Section III.A to compute $\GstukR$;
%\STATE $\forall\;k$, find $\UDk$ and $\URk$ with \eqref{eq:UDkcondition} and \eqref{eq:URkcondition}, respectively;
\STATE Find $\muU$ satisfying $\PU(\muU)=\Ptot$;
\STATE Find $\muL$ satisfying $\PL(\muL)=\Ptot$;
\STATE For each $\mu$ in $S = \{\muL + \frac{n}{N_s}(\muU-\muL), n = 0,1,\cdots,N_s-1\}$,
       use {\bf Algorithm} \ref{alg:LagMax} to find the optimum RA variables to \eqref{eq:LagMax},
       and compute $P_x(\mu)$;
\STATE Initialize $\mu$ with the $\mu\in{S}$ satisfying $P_x(\mu)\leq\Ptot$ and having the minimum $\Ptot-P_x(\mu)$;
\REPEAT
       \STATE Update $\mu$ with $[\mu - \delta(\Ptot-P_x(\mu))]^+$;
       \STATE Use {\bf Algorithm} \ref{alg:LagMax} to find the optimum solution for \eqref{eq:LagMax};
       \STATE Compute $P_x(\mu)$;
\UNTIL{$\mu>0$ and $0\leq\Ptot-P_x(\mu)<\epsilon$}
\STATE The optimum solution for \eqref{eq:LagMax} found in the last iteration
       corresponds to the optimum sum power and mode for each subcarrier in the second step.
%\STATE Use the algorithm described in Section III.A to allocate the optimum sum power for each subcarrier to
%       determine optimum $\Rsetuk$ and $\{\Psuk,\Piuk,\forall\;\Relayi\in\Rsetuk\}$ for relay-aided subcarriers,
%       and the optimum $\PsBuk$ and $\PsRuk$ for subcarrriers in direct mode.
}
\end{algorithmic}
\end{algorithm}

%------------------------------------------------------------------------------
%----------------------------- outer-loop -------------------------------------
%------------------------------------------------------------------------------

The procedure of finding the optimum RA variables to \eqref{eq:LagMax}
is summarized in {\bf Algorithm} \ref{alg:LagMax}.
Next, we address the problem of finding $\muo$ in the outer loop iteratively.
To speed up the convergence rate,
$\mu$ is first initialized by a value close to $\muo$.
To this end, we define
\begin{align}
\PL(\mu)   =&  \sum_{k=1}^K\left[\frac{\min_u\{\wu\}}{\mu} - \valUk \right]^+ \\
\PU(\mu)   =&  \sum_{k=1}^K\left[\frac{2\max_u\{\wu\}}{\mu} - \valLk \right]^+,
\end{align}
where
\begin{align}
    \valUk =& \max\left\{\max_{u\notin\UDk}\frac{1}{\GstukR}, \max_{u\notin\URk}\frac{2}{\Gstuk}\right\} \\
    \valLk =& \min\left\{\min_{u\notin\UDk}\frac{1}{\GstukR}, \min_{u\notin\URk}\frac{2}{\Gstuk}\right\}.
\end{align}

It can be readily shown that $\PL(\muo){\leq}P_x(\muo){\leq}\PU(\muo)$ always holds.
Suppose $\muU$ and $\muL$ satisfy $\PU(\muU)=\Ptot$ and $\PL(\muL)=\Ptot$, respectively.
Since $\PU(\muU)=\Ptot=P_x(\muo){\leq}\PU(\muo)$ and $\PU(\mu)$ is a non-increasing function of $\mu$,
$\muU\geq\muo$. Similarly, we can show $\muo\geq\muL$,
thus $\muo$ is confined in the interval $[\muL,\muU]$.
To initialize $\mu$ with a value close to $\muo$,
this interval is sampled uniformly at $N_s$ points to build the set
$S = \{\muL + \frac{n}{N_s}(\muU-\muL), n = 0,1,\cdots,N_s-1\}$.
Given each $\mu$ in $S$, \eqref{eq:LagMax} is solved and $P_x(\mu)$ computed,
then $\mu$ is initialized with the $\mu\in{S}$
satisfying $P_x(\mu)\leq\Ptot$ and having the minimum $\Ptot-P_x(\mu)$.

After initialization,
$\mu$ is iteratively updated based on the subgradient method,
i.e., $\mu$ is updated with $[\mu - \delta(\Ptot-P_x(\mu))]^+$
where $\delta$ is a sufficiently small value to guarantee convergence,
until $\muo$ is found when $\Ptot=P_x(\mu)$ and $\mu>0$.
The RA algorithm in the second step is summarized in {\bf Algorithm} \ref{alg:step-2}.
Due to numerical issues,
we regard the KKT conditions to be satisfied
when $\mu>0$ and $0\leq\Ptot-P_x(\mu)\leq\epsilon$, where $\epsilon$ is a very small positive value.

\section{The optimum RA in the second step for a special case}\label{sec:PtotHigh}

We will show the optimum RA in the second step can be easily derived
when $\Ptot$ is very high so that the associated $\muU$ satisfies
\begin{align}
   \forall\; k, \; \frac{\min_u\wu}{\muU}   \gg& \max_u\{\max\{\frac{1}{\GstukR}, \frac{1}{\Gstuk}\}\},    \label{eq:BigPcond1}\\
   \forall\; k, \; \frac{\min_u\wu}{\muU}   \gg& \max_u\{\max\{\GstukR, \Gstuk \}\}.                        \label{eq:BigPcond2}
\end{align}

{Since \eqref{eq:BigPcond1} holds and $\muo\leq\muU$, the optimum sum power $P_k$ for subcarrier $k$ satisfies
\begin{align}
   \forall\,k,\, P_k \approx \left\{
                           \begin{array}{ll}
                               \frac{w_{\uk}}{\muo}    &   \mathrm{if}\, \ik = 1   \\
                               \frac{2w_{\uk}}{\muo}   &   \mathrm{if}\, \ik = 2,
                           \end{array}
                  \right. \label{eq:apprPk}
\end{align}
according to \eqref{eq:optPukRD}, where $(\uk,\ik)$ corresponds to
the optimum destination and mode for subcarrier $k$ found by the RA algorithm in the second step.
From \eqref{eq:BigPcond1}, \eqref{eq:BigPcond2}, and \eqref{eq:apprPk},
the conditions
\begin{align}
   \forall\,k,u,             P_k\Gstuk\gg1,   &  P_k\gg\Gstuk, \label{eq:BigPcond3}\\
   \forall\,k,u,   P_k\GstukR\gg1,  &  P_k\gg\GstukR, \label{eq:BigPcond4}
\end{align}
hold. This means that in the high power regime,
even though the optimum sum power for each subcarrier is not precisely known,
it is for sure that $\forall\;k,u$, the following approximations hold:
\begin{align}
  \wu\RukD(P_k) \approx &  2\wu\big(\ln(P_k)+\ln\frac{\Gstuk}{2}\big) \approx  2\wu\ln(P_k),  \label{eq:BigP-apprRukD}\\
  \wu\RukR(P_k) \approx &   \wu\big(\ln(P_k) + \ln\GstukR\big) \approx \wu\ln(P_k).              \label{eq:BigP-apprRukR}
\end{align}

By comparing \eqref{eq:BigP-apprRukD} and \eqref{eq:BigP-apprRukR},
it can easily be seen that the direct mode for each subcarrier $k$ is
the optimum for maximizing the WSR, independently of the destination assigned.
All this confirms the former interpretation that,
for the high power regime, increasing the number of channel uses
per time unit is more beneficial than increasing the received power.}
If equal weights are used for all destinations,
the optimum destination for subcarrier $k$ is the one with the maximum $\Gstuk$.
Otherwise, the optimum destination for all subcarriers is the same
and given by the one with the maximum weight $\wu$.
In either case,
the optimum sum power to subcarrier $k$ is
$P_k \approx \frac{\Ptot}{K}$ in order to satisfy the KKT conditions according to \eqref{eq:apprPk}.

\section{Numerical experiments}

%\subsection{An example system setup and a reference protocol}

In numerical experiments, we consider a multiple DF relays aided OFDMA downlink
transmission system with $N=4$ relays. %$K=64$, $N=4$, and $U=6$.
The source is allocated at the origin, and
the relays $\mathrm{r}_1$, $\mathrm{r}_2$, $\mathrm{r}_3$, and $\mathrm{r}_4$
are located at coordinates $(-15, -5)$, $(-5, -5)$, $(5, -5)$, and $(15, -5)$, respectively.
Note that all the aforementioned coordinate-related values have the unit of meter.
The random channel between every two of the source, the relays, and the destinations
is generated based on a $6$-tap delay line model,
with the $i$-th tap's amplitude as a circular Gaussian random variable of zero mean
and variance as $\sigma_i^2 = \sigma_0^2{e}^{-3i}$ $(i=0,1,\cdots,5)$.
Moreover, the average received power attenuation is equal to $\alpha{d}^{-3}$ at a distance of $d$.
Note that $10\log_{10}(\alpha)$, representing the log-normal shadowing effect,
is Gaussian distributed with zero mean and variance equal to $0$ dB.
This means that the transmit power is attenuated by $30$ dB in average at a distance of $10$ m.

{To illustrate the effectiveness of the optimum RA found for the proposed protocol, we compare it to the optimum RA for a reference protocol.
Both protocols are the same except that with the reference protocol,
at every subcarrier in the direct mode the source only transmits
in the broadcasting slot (and remains idle during the relaying slot).
Interestingly, when the destinations use {\it equal} weights,
the optimum RA for the reference protocol
can easily be derived on a per subcarrier basis. For subcarrier $k$, the optimum destination $\uk$ is the one with the maximum $G_u(k)$
where $G_u(k) = \max\{\GstukR,\Gstuk\}$,
i.e., the relay-aided mode is optimum only if $G_{\uk,1}(k)>G_{{\rm s},\uk}(k)$.
The optimum per subcarrier sum power can be calculated according to
water filling of the system sum power $\Ptot$ to $K$ parallel channels
with normalized gain $G_{\uk}(k)$ for the $k$-th channel.
It is important to note that for the reference protocol,
the above RA might {\it not} be optimum when destinations use {\it unequal} weights.
In this case, the RA needs to be optimized jointly over all subcarriers
with the same two-step optimization strategy as proposed in Section III,
which is not developed here due to space limitation.
On the contrary, the RA for the proposed protocol {\it always} needs to be jointly optimized over all subcarriers.
}

\begin{figure}[!t]
  \centering
     \includegraphics[width=3in]{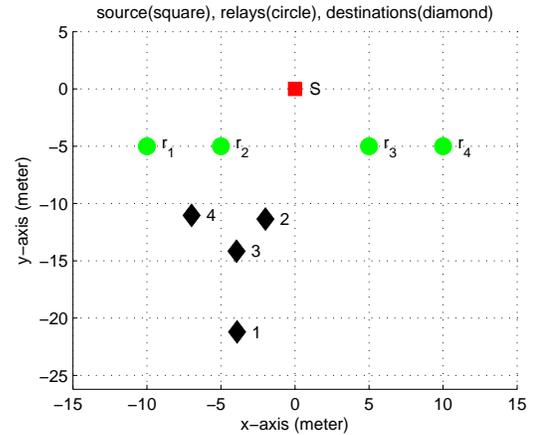}
  \caption{The OFDMA system considered in numerical experiments.}
  \label{fig:Systeup}
\end{figure}

\subsection{Results for a random realization of channels}

\begin{figure}[!t]
  \centering
  \subfigure[]{
     \includegraphics[width=3.5in]{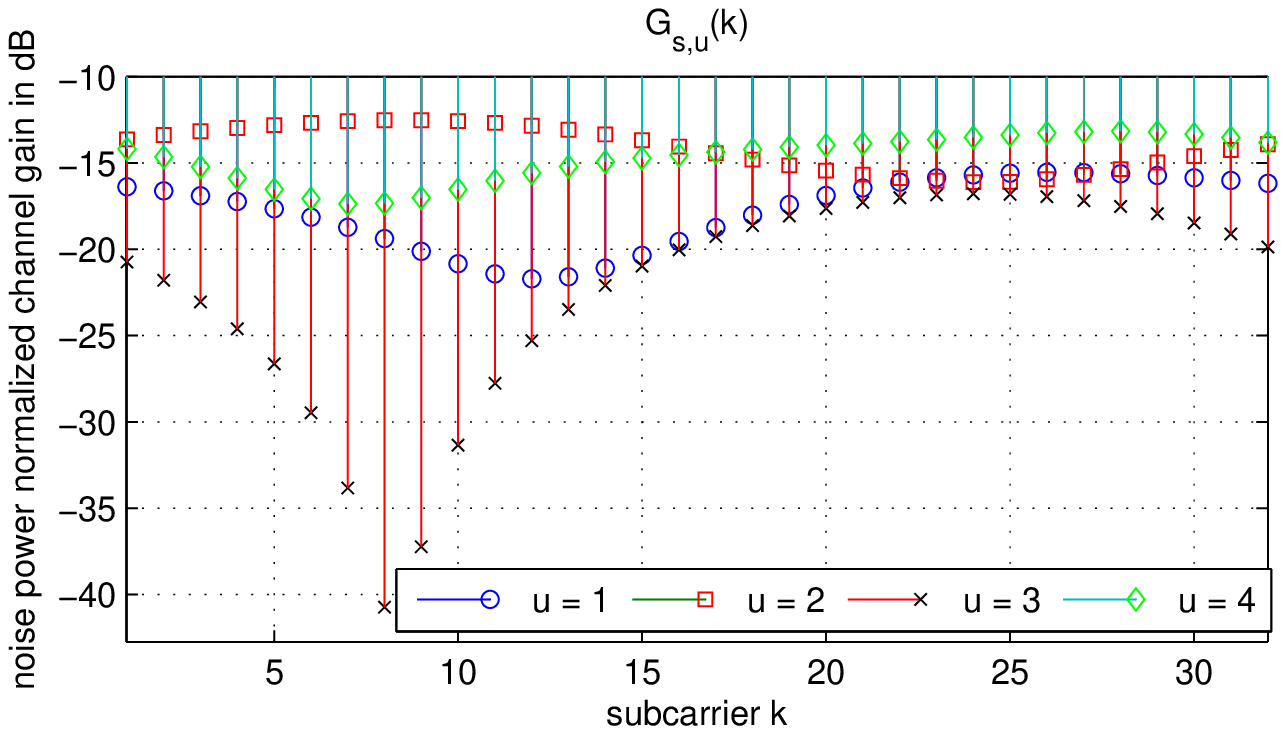}}
  \subfigure[]{
     \includegraphics[width=3.5in]{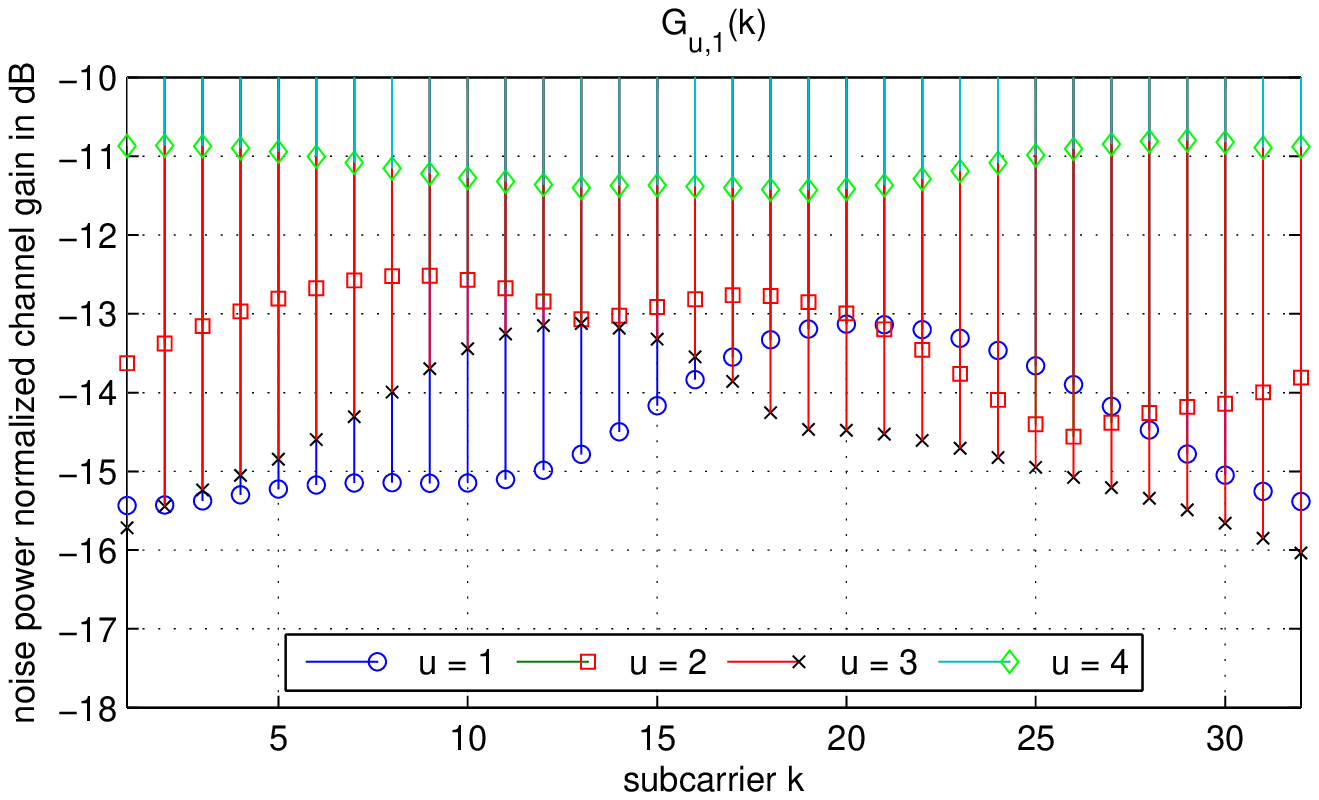}}
  \caption{(a) $\Gstuk$ and (b) $\GstukR$ computed by the proposed RA algorithm in the first step,
               at each subcarrier $k$ for every destination.
           }
  \label{fig:Gains}
\end{figure}

For clarity of presentation, we first assume $K=32$ subcarriers are used,
$\NVar = -30$ dBW, $U=4$ destinations exist with equal weights,
$N_s=100$, $\delta=10^{-3}(\muU-\muL)$, and $\epsilon=0.1$.
For the randomly generated destination coordinates shown in Figure \ref{fig:Systeup},
we have produced a random realization of all channels (not shown here due to space limitation),
over which the optimum RAs for the proposed and reference protocols are evaluated.
For every destination $u$,
the $\GstukR$ computed by the proposed RA algorithm in the first step,
and $\Gstuk$ are shown in Figure \ref{fig:Gains} for every subcarrier $k$ and destination.

%======================================== 35dBW =======================================
\begin{figure}
  \centering
  \subfigure[]{
    \includegraphics[width=3.5in]{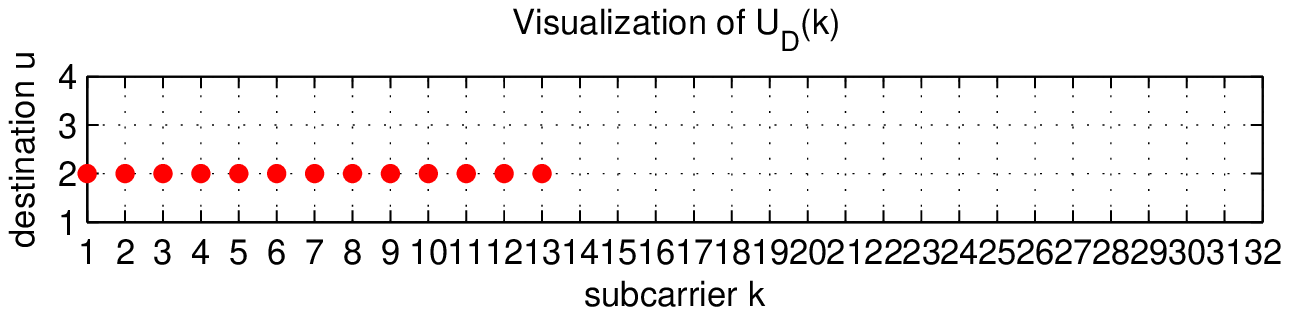}}
  \subfigure[]{
    \includegraphics[width=3.5in]{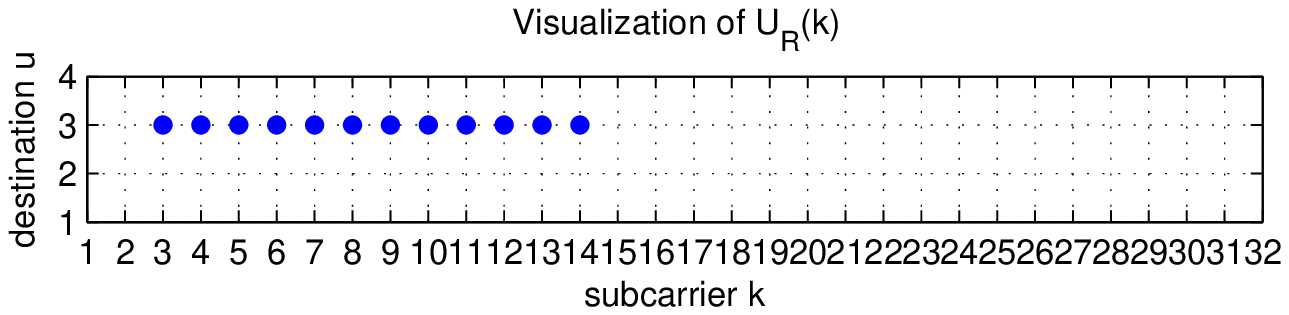}}
  \caption{When $\Ptot=35$ dBW,
           (a) the visualization of $\UDk$, $\forall\;k$,
           where a dot marked at the coordinate ($k$,$u$) indicates destination $u\in\UDk$, otherwise $u\notin\UDk$.
           (b) the visualization of $\URk$, $\forall\;k$,
           where a dot marked at the coordinate ($k$,$u$) indicates destination $u\in\URk$, otherwise $u\notin\URk$.
           }

  \label{fig:URk-30dBW}
\end{figure}

\begin{figure}
  \centering
  \subfigure[]{
     \includegraphics[width=3.5in,height=1.5in]{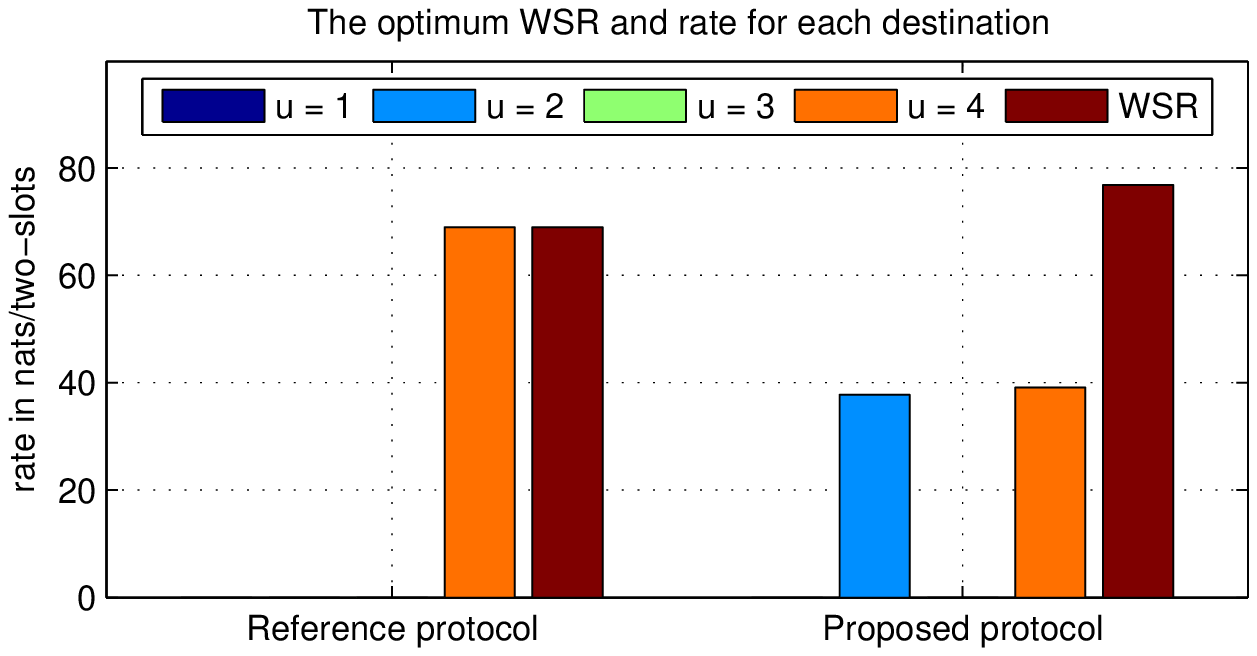}}
  \subfigure[]{
     \includegraphics[width=3.5in]{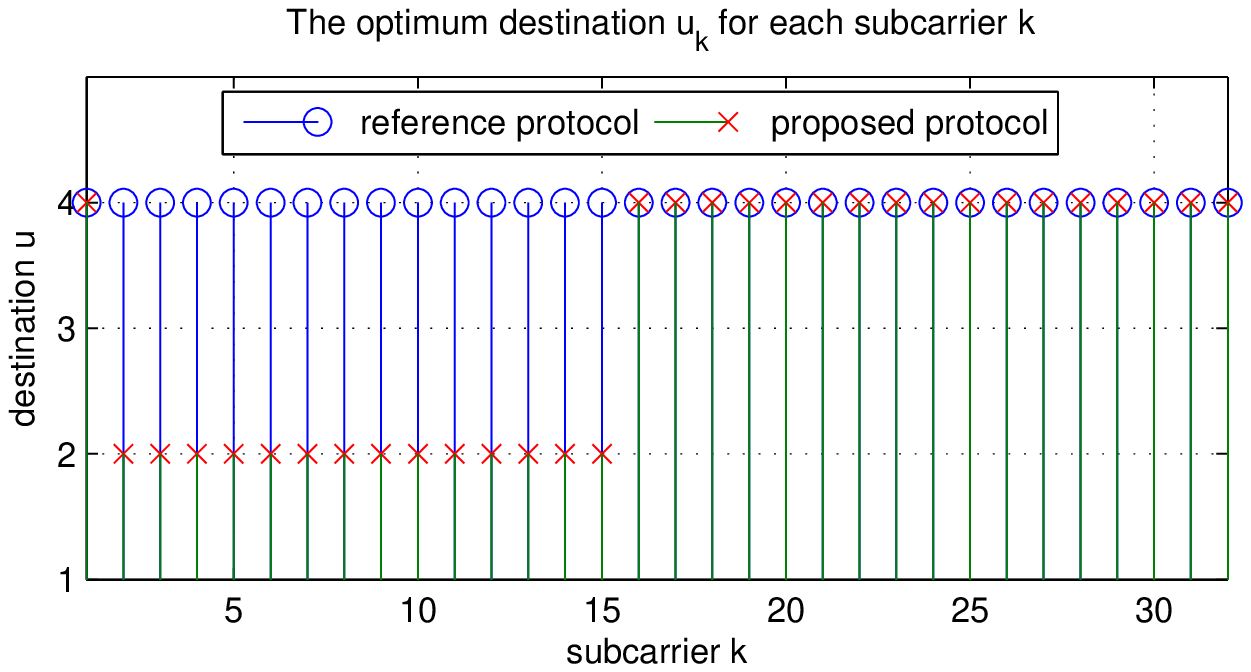}}
  \subfigure[]{
     \includegraphics[width=3.5in]{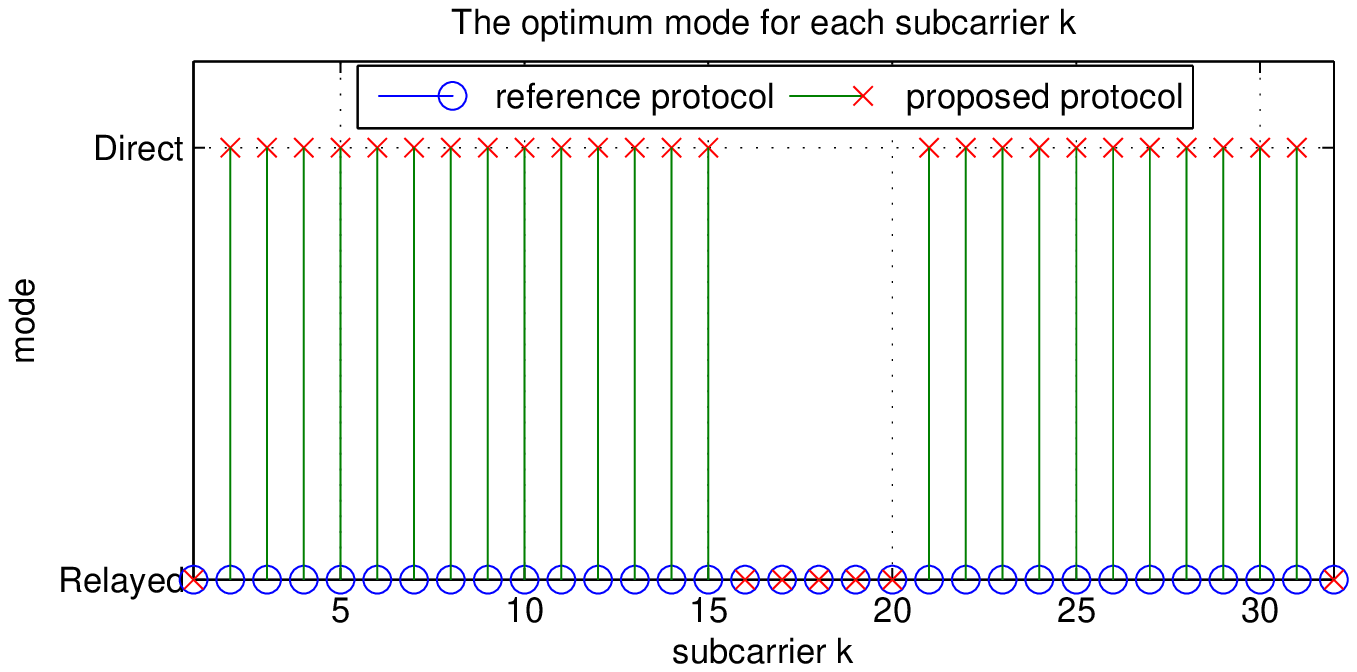}}
  \subfigure[]{
     \includegraphics[width=3.5in]{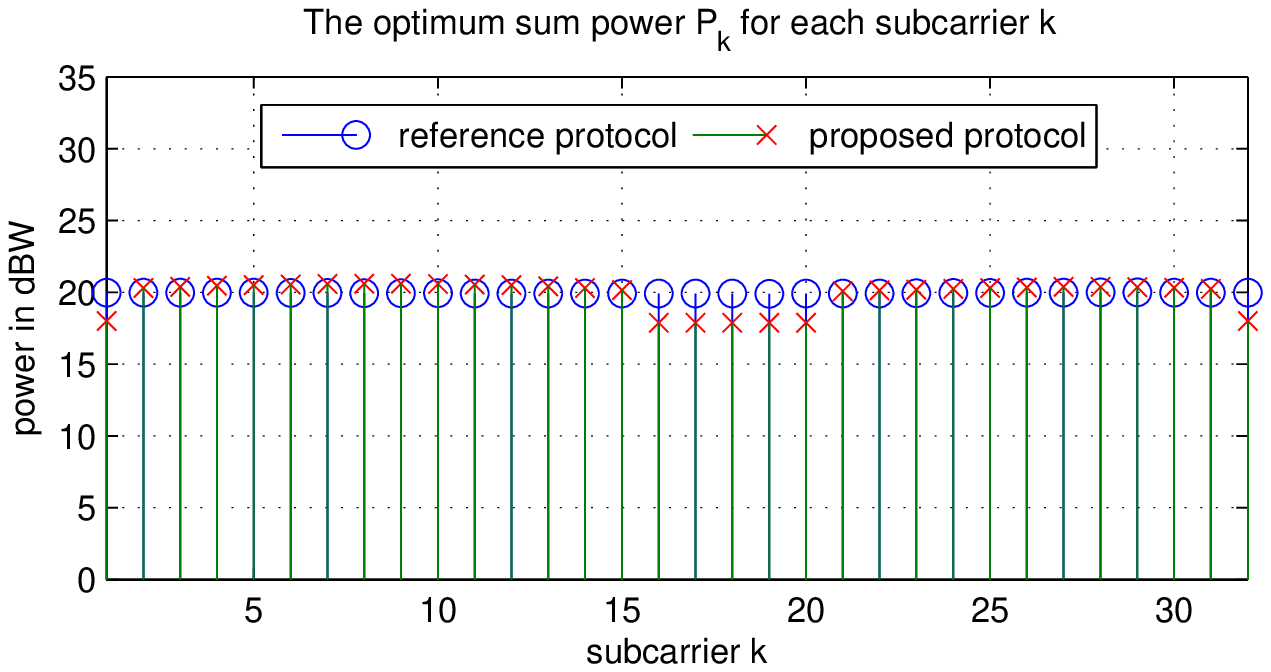}}
  \caption{For the reference and proposed protocols, the optimum RA found in the second step
           for each subcarrier $k$ when $\Ptot=35$ dBW.}
  \label{fig:OptRA-35dBW}
\end{figure}

\begin{figure}
  \centering
  \subfigure[]{
     \includegraphics[width=3.5in]{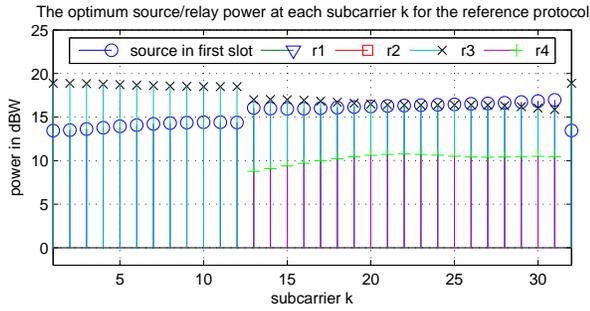}}
  \subfigure[]{
     \includegraphics[width=3.5in]{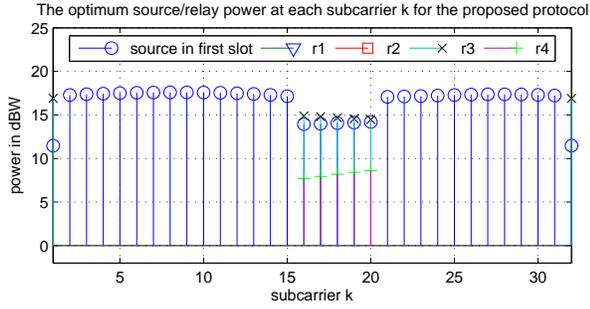}}
  \caption{The optimum source and relay power in the broadcasting slot for the two protocols when $\Ptot=35$ dBW.
           }
  \label{fig:OptPSR-35dBW}
\end{figure}

First, we set $\Ptot=35$ dBW which corresponds to the low power regime,
since if $\Ptot$ is uniformly allocated to the subcarriers of the source,
the average SNR is $2.17$, $5.80$, $-2.38$, and $5.23$ dB at four destinations, respectively.
The corresponding $\UDk$ and $\URk$ are found and visualized in Figure \ref{fig:URk-30dBW} for every subcarrier.
The optimum RA and the corresponding WSR
have been evaluated for the proposed protocol with the RA algorithm developed in Section III,
and for the reference protocol with the aforementioned method.
The results are shown in Figure \ref{fig:OptRA-35dBW}.
{We can see that for the reference protocol,
the optimum destination and mode at every subcarrier $k$ are $\uk=4$ with
the maximum $G_u(k) = \max\{\GstukR,\Gstuk\}$ and the relay-aided mode, respectively.
The uniform sum power allocation to all subcarriers is optimum.
The reason is that $\GstukR$ for destination $4$ varies slightly over all subcarriers.
%therefore the uniform sum power allocation is optimum because of
%the aforementioned water filling to find the optimum subcarrier sum power for the reference protocol
%(refer to Section $5.4.6$ in \cite{Fund-WCOM} for more details).
{\it
It is very interesting to note that,
compared to the reference protocol,
assigning a few subcarriers to destination $2$ or $4$ in the direct mode
with a higher per subcarrier sum power,
leads to a higher optimum WSR for the proposed protocol,
even though at those subcarriers $\Gstuk$ for destination $2$ or $4$ is smaller than $\GstukR$ for destination $4$.}
For both protocols, the optimum $\PsBuk$ and $\{\Piuk | \forall\;\Relayi\}$
computed by the RA algorithm in the first step are shown in Figure \ref{fig:OptPSR-35dBW}.
It is shown that $r_3$ and $r_4$ assist relaying simultaneously at a few subcarriers
for both protocols.
}

%======================================== 60dBW =======================================

\begin{figure}
  \centering
  \includegraphics[width=3.5in]{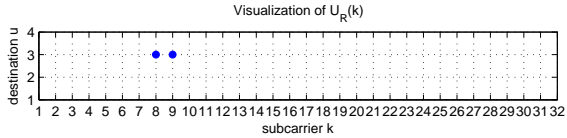}
  \caption{The visualization of $\URk$ for each subcarrier $k$ when $\Ptot=60$ dBW,
           where a dot marked at the coordinate ($k$,$u$) means that destination $u\in\URk$, otherwise $u\notin\URk$.}
  \label{fig:URk-c3-60dBW}
\end{figure}

Next, we set $\Ptot=60$ dBW, which satisfies the conditions \eqref{eq:BigPcond1} and \eqref{eq:BigPcond2}.
For each subcarrier $k$, the $\UDk$ is the same as when $\Ptot=35$ dBW,
since it is determined independently of $\Ptot$ according to  the conditions in \eqref{eq:UDkcondition}.
The $\URk$ is visualized in Figure \ref{fig:URk-c3-60dBW} for every subcarrier.
Now, $\URk$ contains none of the destinations at most of the subcarriers,
since $\Ptot$ is so high that the conditions in \eqref{eq:URkcondition} are not satisfied
by any destination at many subcarriers.
The optimum RAs and WSRs have been evaluated for the two protocols and shown in Figure \ref{fig:OptRA-60dBW}.
{It is shown that the optimum RA for the proposed protocol leads to
a much higher WSR compared to that for the reference protocol.
For the reference protocol, the optimum mode and destination at each subcarrier
are the same as when $\Ptot=35$ dBW.
This is because when the destinations have equal weights,
the optimum mode and destination for the reference protocol
only depends on the $\Gstuk$ and $\GstukR$ of the destinations.
The uniform sum power allocation to all subcarriers is still optimum, as explained for the case when $\Ptot=35$ dBW.
For the proposed protocol,
the optimum destination and mode for every subcarrier $k$ are the destination with
the maximum $\Gstuk$ and the direct mode, respectively,
and the uniform sum power allocation is also optimum.
For the reference protocol, the optimum $\PsBuk$ and $\{\Piuk | \forall\;\Relayi\}$
computed by the RA algorithm in the first step are shown in Figure \ref{fig:OptPSR-60dBW}.
It is shown that $r_3$ and $r_4$ assist relaying simultaneously at a few subcarriers.
Note that for the proposed protocol, the optimum sum power at subcarrier $k$ is equally
allocated to $\PsBuk$ and $\PsRuk$, which is not shown here.

\begin{figure}
  \centering
  \subfigure[]{
     \includegraphics[width=3.5in,height=1.5in]{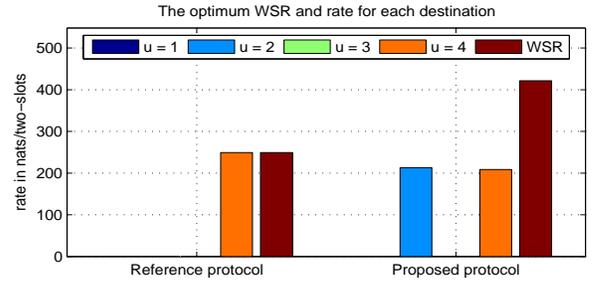}}
  \subfigure[]{
     \includegraphics[width=3.5in]{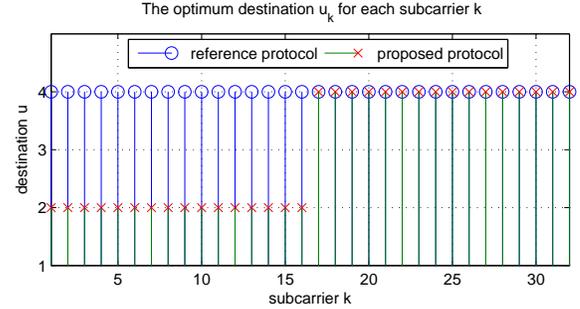}}
  \subfigure[]{
     \includegraphics[width=3.5in]{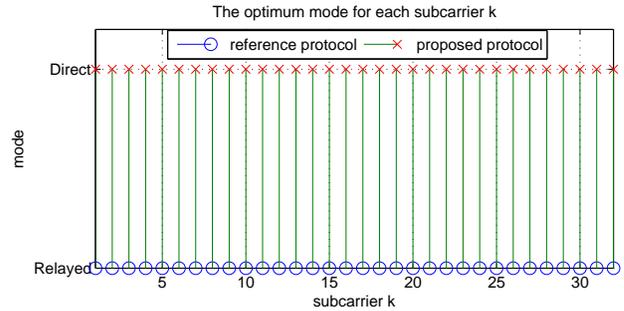}}
  \subfigure[]{
     \includegraphics[width=3.5in]{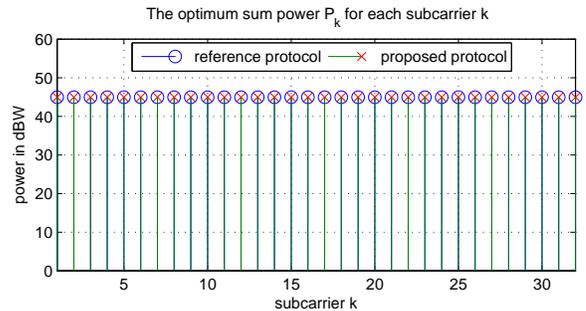}}
  \caption{For the reference and proposed protocols, the optimum RA found in the second step
           for each subcarrier $k$ when $\Ptot=60$ dBW and destinations have equal weights.}
  \label{fig:OptRA-60dBW}
\end{figure}

\begin{figure}[!b]
  \centering
     \includegraphics[width=3.5in]{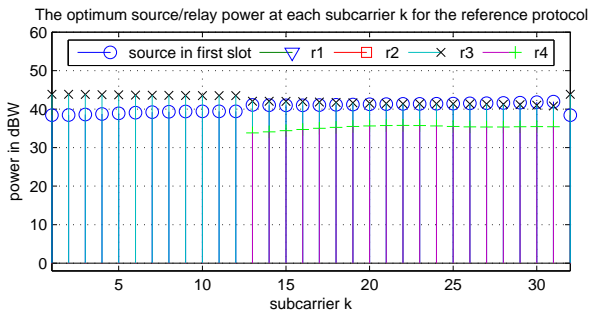}
  \caption{The optimum source and relay power in the broadcasting slot for the reference protocol when $\Ptot=60$ dBW.
           Note that for the proposed protocol, the optimum sum power at each subcarrier $k$ is equally allocated to $\PsBuk$ and $\PsRuk$,
           which is not shown here.}
  \label{fig:OptPSR-60dBW}
\end{figure}

%================================ unequal weights ==========================================
\begin{figure}
  \centering
  \subfigure[]{
     \includegraphics[width=3.5in]{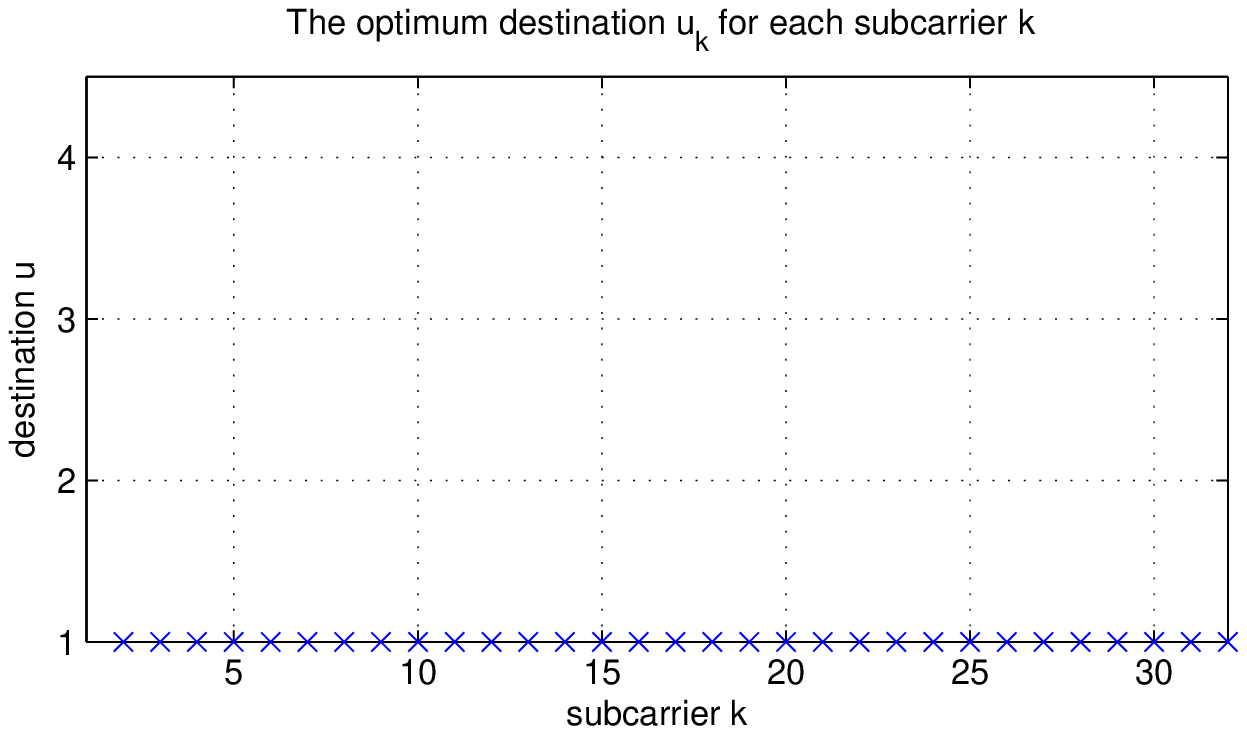}}
  \subfigure[]{
     \includegraphics[width=3.5in]{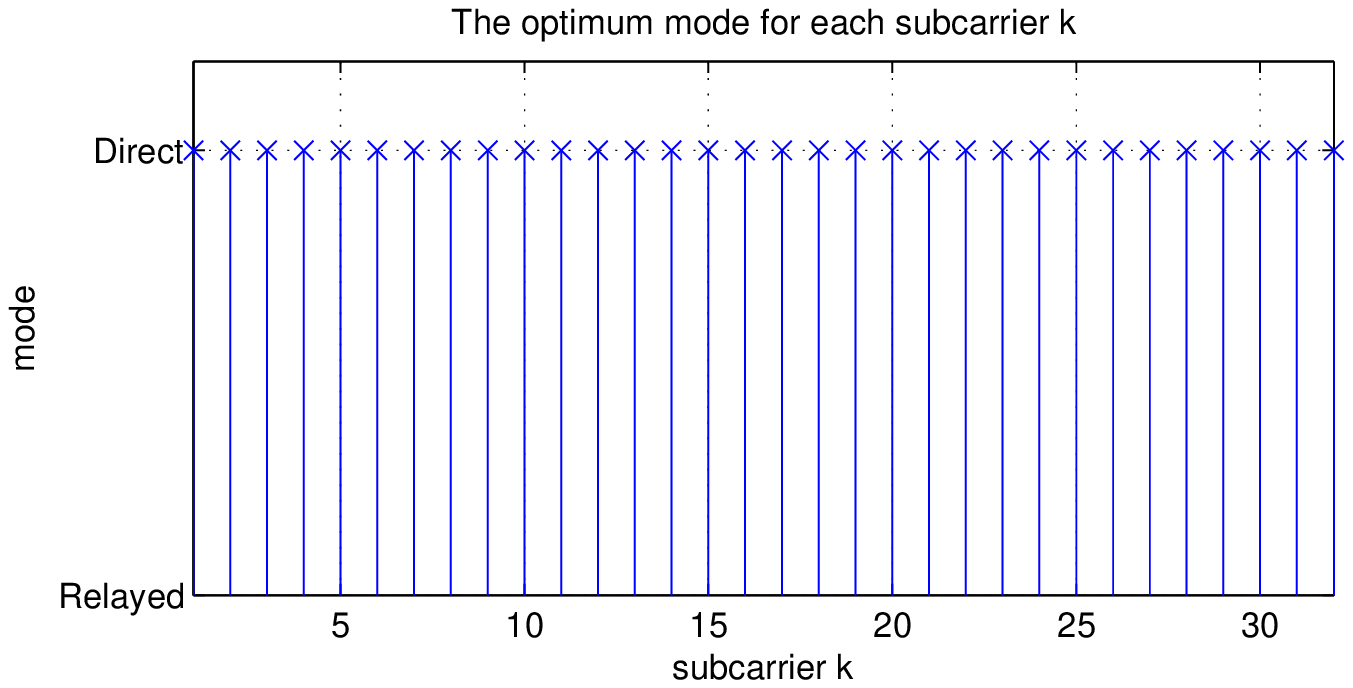}}
  \subfigure[]{
     \includegraphics[width=3.5in]{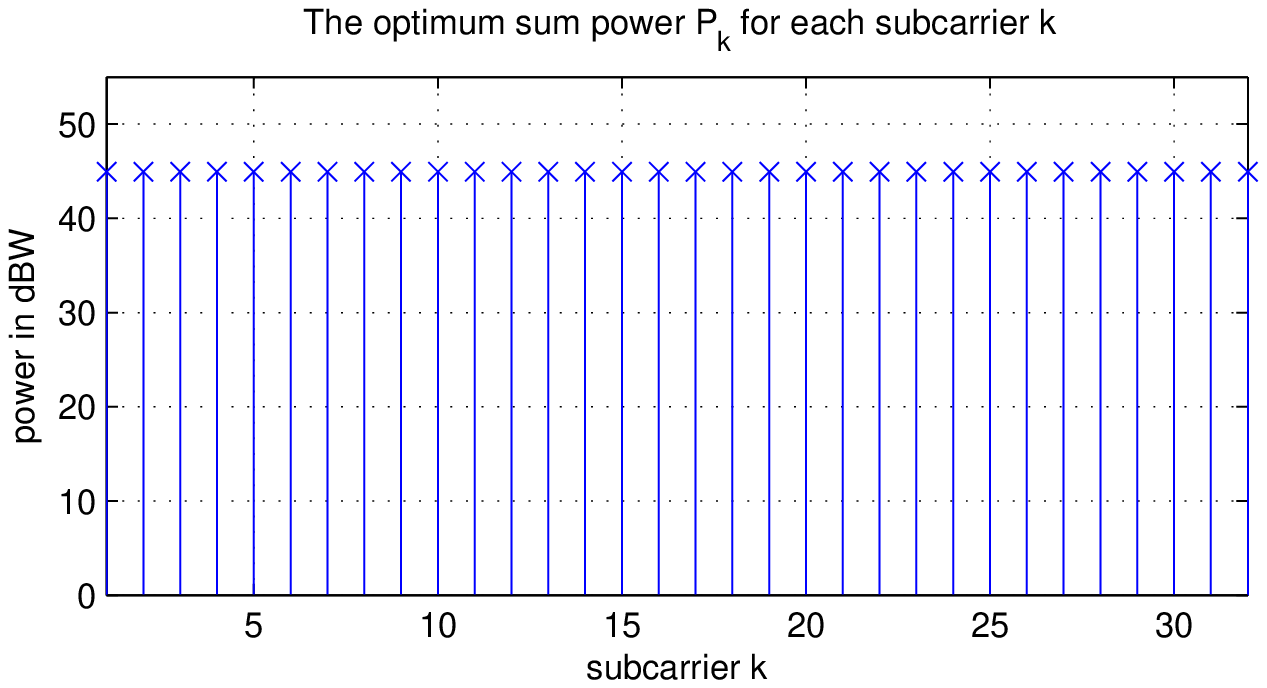}}
  \caption{For the proposed protocol, the optimum RA found in the second step
           for each subcarrier $k$ when $\Ptot=60$ dBW and destination $4$ has the maximum weight.}
  \label{fig:OptRA-60dBW-uneqw}
\end{figure}

For the considered channel realization,
the above observations illustrate the effectiveness of the proposed protocol
and algorithm to adapt the RA dynamically for WSR maximization. They also support
the analysis of Section IV for the high power regime.
When $\Ptot=60$ dBW and the weights for destination $1$, $2$, $3$, and $4$
are $0.4$, $0.2$, $0.2$, $0.2$, respectively, the optimum RA for the proposed
protocol has also been evaluated and shown in Figure \ref{fig:OptRA-60dBW-uneqw}.
We can see that destination $1$ with the maximum weight and the direct mode
is optimum for each subcarrier.
The uniform sum power allocation is optimum.
These observations illustrate the effectiveness of the analysis of Section IV
for unequal destination weights in the high power regime.}

\subsection{Results for a set of random realizations of channels}

\begin{figure}
  \centering
  \subfigure[when $\Ptot=35$ dBW]{
     \includegraphics[width=3.5in]{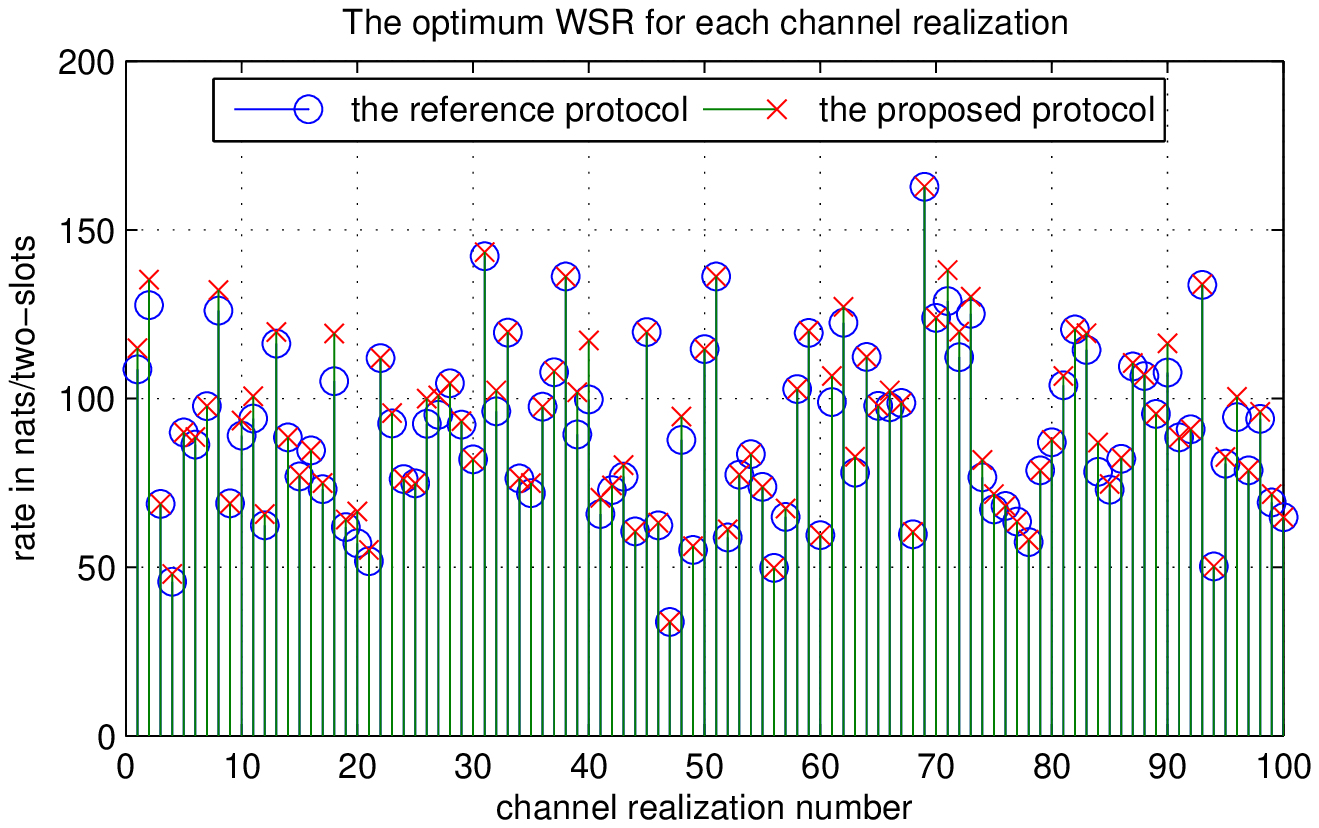}}
  \subfigure[when $\Ptot=60$ dBW]{
     \includegraphics[width=3.5in]{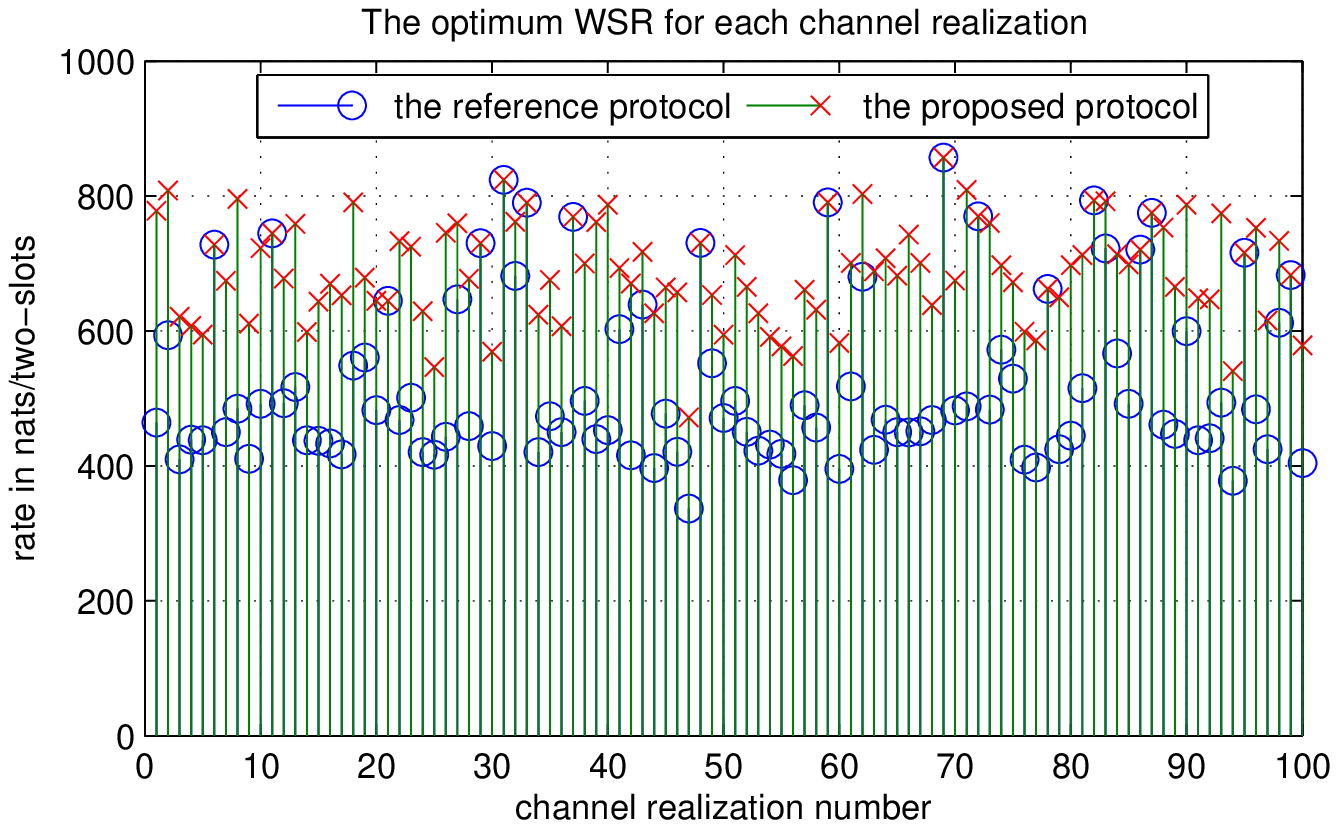}}
  \subfigure[when $\Ptot=35$ or $60$ dBW]{
     \includegraphics[width=3.5in]{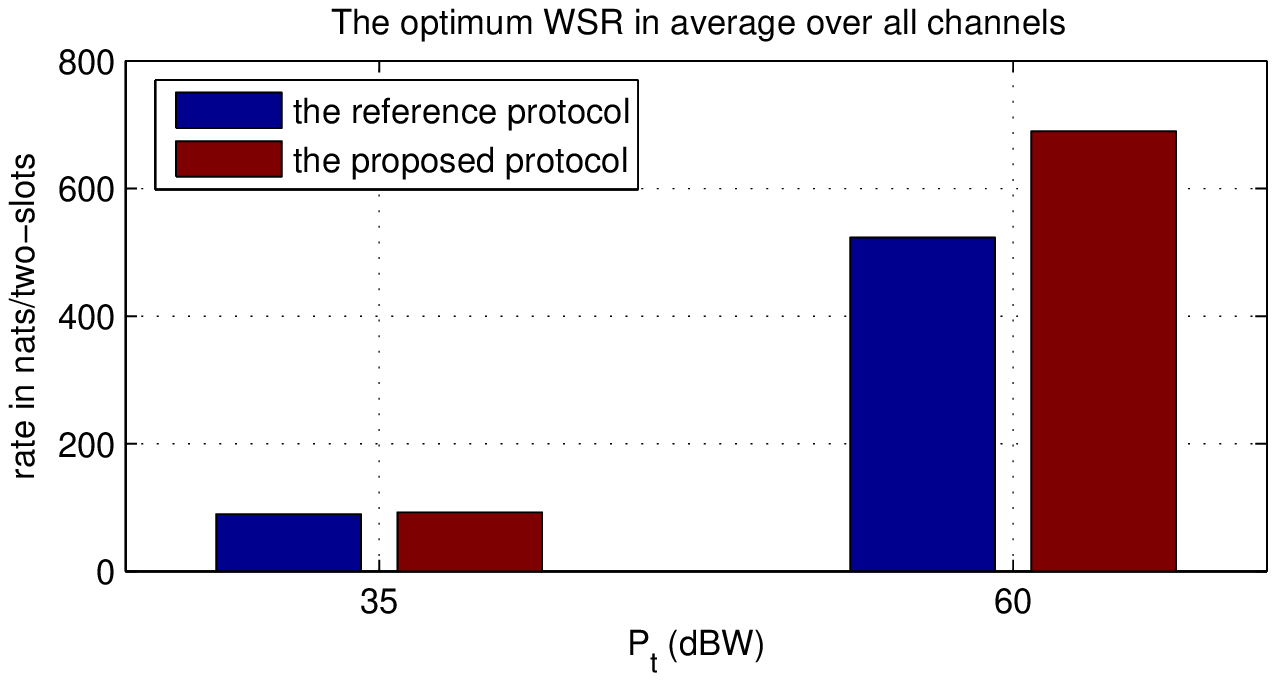}}
  \caption{The optimum WSRs for the first $100$ channel realizations for the reference and proposed protocols,
           and the average WSR for all channel realizations. }
  \label{fig:allWSR}
\end{figure}

For the comprehensive illustration of scenarios close to reality,
we have generated $1000$ random channel realizations.
For each realization, the optimum RAs and the corresponding WSRs for the reference and proposed protocols are evaluated.
We assume $K=64$ subcarriers are used, $\NVar = -30$ dBW,
$U=8$ destinations exist with equal weights,
$N_s=100$, $\delta=10^{-3}(\muU-\muL)$, and $\epsilon=0.1$.
To produce each channel realization,
the destination coordinates are first generated
in the region $\{(x,y)|-10\leq{x}\leq10,-30\leq{y}\leq-10\}$
uniformly and independently of each other,
then the random channels are produced.

For better clarity, the optimum WSRs for only the first $100$ channel realizations,
and the average WSR for all channel realizations are shown
in Figure \ref{fig:allWSR} for both protocols when $\Ptot$ is $35$ and $60$ dBW, respectively.
It is found that for every channel realization,
the optimum WSR of the proposed protocol is equal to or greater than that for the reference protocol.
The reason is that the set of feasible RAs searched for the optimum RA of the reference protocol,
is a subset of that of the proposed protocol,
since extra constraints are imposed on the RA optimization for the reference protocol,
i.e. $\PsRuk$ must be zero for every subcarrier in the direct mode.
Compared to the reference protocol, the optimum RA for the proposed protocol
leads to either the same or a slightly greater optimum WSR for each channel realization when $\Ptot=35$ dBW.
However, a much higher optimum WSR results from the optimum RA
and the proposed protocol for the majority of channel realizations when $\Ptot=60$ dBW.
As a result, the average optimum WSRs for the two protocols are very close when $\Ptot=35$ dBW,
while the proposed protocol and RA result in a much higher average optimum WSR
than the reference protocol when $\Ptot=60$ dBW.
As already said, this is due to the better exploitation
of channel uses by the direct mode of the proposed protocol
in the high power regime.

\begin{figure}
  \centering
  \subfigure[]{
     \includegraphics[width=3.5in]{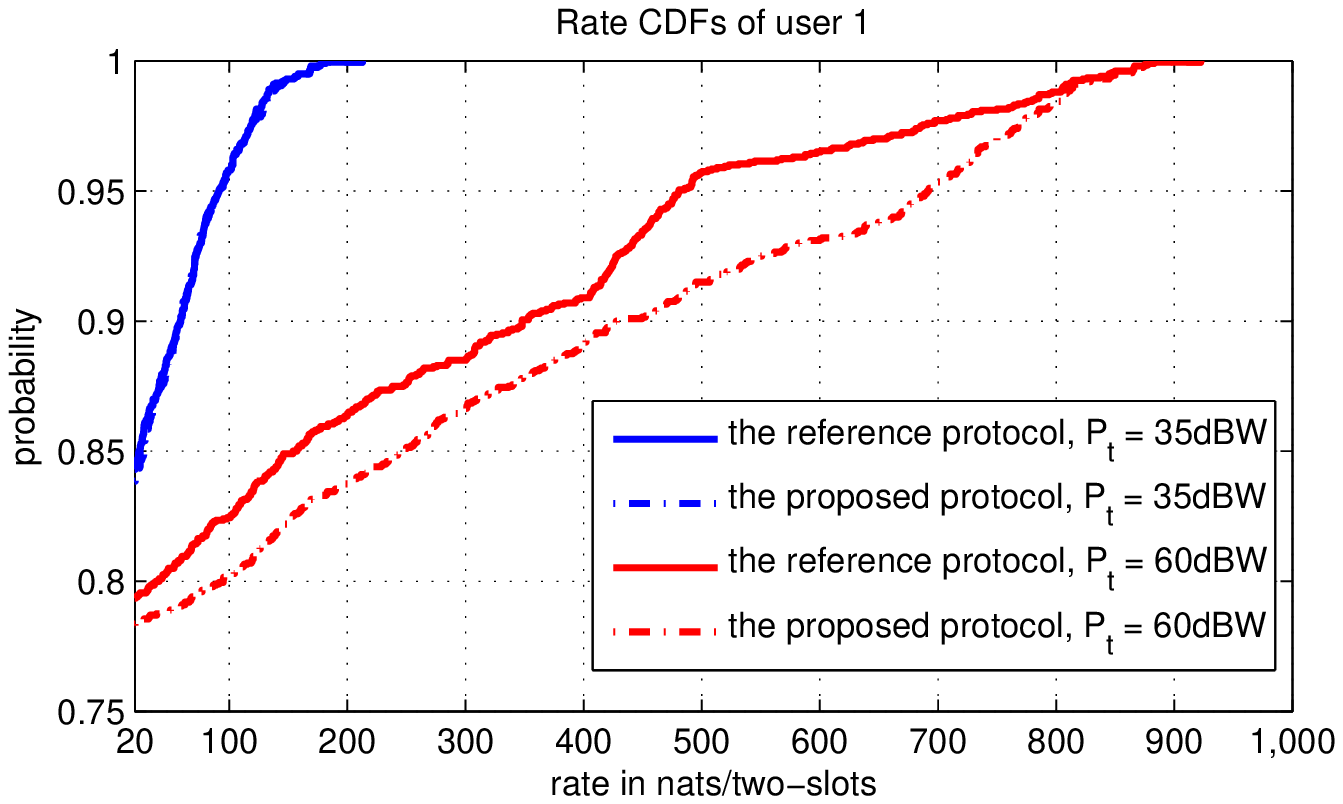}}
  \subfigure[]{
     \includegraphics[width=3.5in]{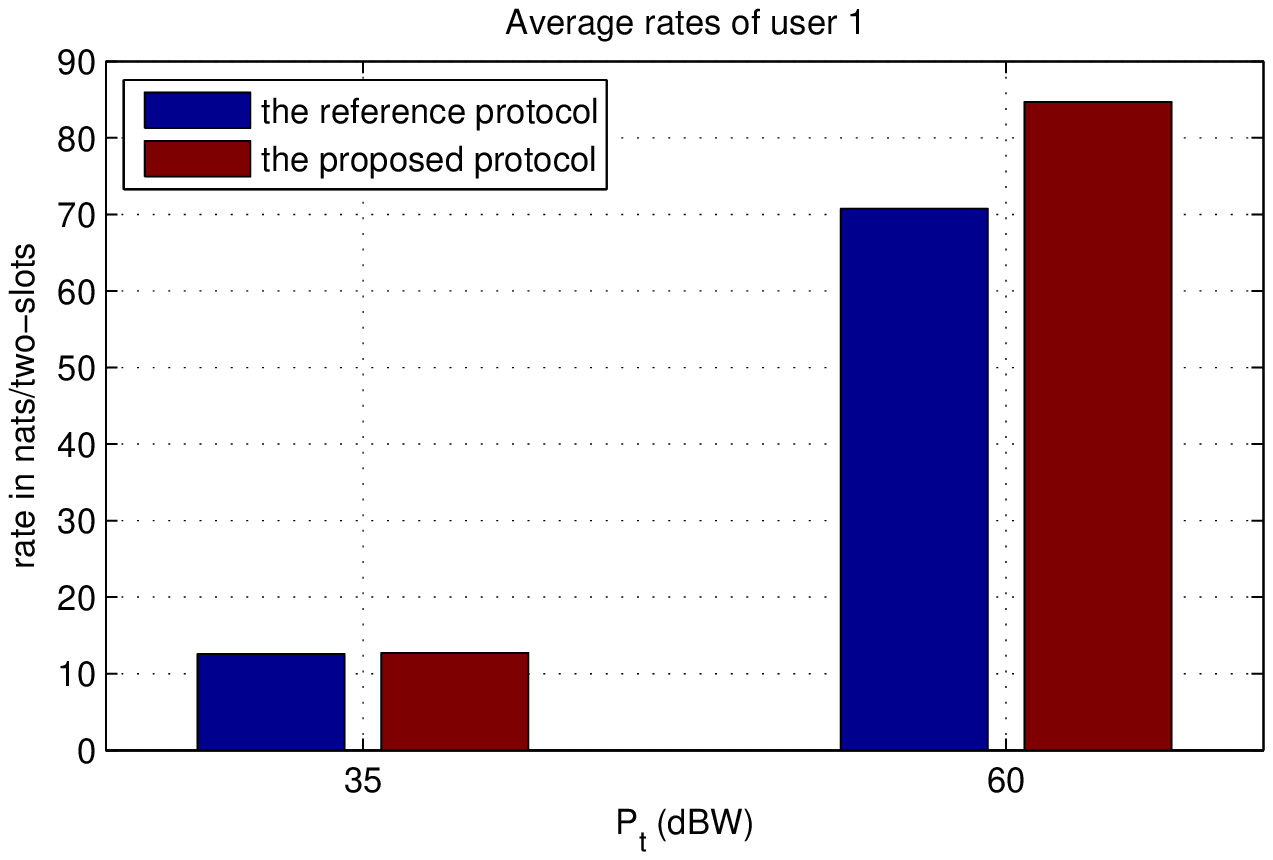}}
  \caption{User $1$'s rate CDFs and average rates for the reference and proposed protocols.}
  \label{fig:CDF}
\end{figure}

For both protocols, the rate cumulative distribution functions (CDF) and average rates of user $1$
are shown in Figure \ref{fig:CDF} when $\Ptot$ is $35$ and $60$ dBW, respectively.
Note that for the same protocol and $\Ptot$, all other users have
similar rate CDFs and average rates as user $1$, since all users' coordinates are independently
and identically distributed in the same region.
It can be seen that both protocols lead to almost the same rate CDFs for user $1$ when $\Ptot=35$ dBW.
Nevertheless, using the proposed protocol makes it more likely to transmit with high rate
to user $1$ when $\Ptot=60$ dBW. This means that for a fixed outage probability,
the proposed protocol leads to a higher outage rate for user $1$ than the reference protocol.
As a result, the average rates of user $1$ for both protocols are almost the same when $\Ptot=35$ dBW,
while the proposed protocol leads to a much higher average rate of user $1$ than the reference protocol when $\Ptot=60$ dBW.
These observations illustrate the effectiveness of the proposed protocol
and algorithm for the random channel realizations.

\section{Conclusion}

We have considered the WSR maximized RA problem constrained by a system sum power
in an OFDMA downlink transmission system assisted by multiple DF relays.
In particular, multiple relays may cooperate with the source for every relay-aided transmission.
We have proposed a two-step algorithm to optimize the RA.
Furthermore, we have shown that the optimum RA in the second step
can be easily derived when the system sum power is very high.
The effectiveness of the proposed algorithm has been illustrated by numerical experiments.

\appendices

\section{Maximization of $\RukR$ in the relay-aided mode}

When $\Gstuk\geq\Gstoxk{\SRelayki{N}}$, $\SNRminRsetuk\leq\SNRumrck$ always holds,
thus $\RukR = \ln(1 + \Psuk\cdot\min_{\Relayi\in\Rsetuk}\Gstik)$,
which means that $\RukR$ is equal to the maximum rate allowed
for reliable transmission from the source to all relays in $\Rsetuk$.
In this case, the maximum $\RukR$ is $\RukR(P) = \ln(1 + \Gstoxk{\SRelayki{N}}P)$.
To achieve it, $\Rsetuk$ and $\Psuk$ should be equal to $\{\SRelayki{N}\}$
and $P$, respectively.

Now we consider the case when $\Gstoxk{\SRelayki{N}}>\Gstuk$.
The maximization of $\RukR$ is equivalent to the maximization of $\SNRuk$,
as the minimum of $\SNRumrck$ and $\SNRminRsetuk$.
Note that $\SNRminRsetuk$ depends on $\Psuk$ and $\Rsetuk$,
while besides them $\SNRumrck$ depends on $\{\Piuk | \Relayi\in\Rsetuk\}$ as well.
To maximize $\SNRuk$ given $P$, $\SNRumrck$ can be first maximized
with $\Psuk$ and $\Rsetuk$ fixed, by finding the optimum allocation of
the sum relay power $P-\Psuk$ to $\{\Piuk | \Relayi\in\Rsetuk\}$.
Denoting such maximized $\SNRumrck$ by $\MaxMRCuk$,
the optimum $\{\Psuk,\Rsetuk\}$ that maximizes $\min\{\MaxMRCuk,\SNRminRsetuk\}$ is then found.
Finally, that $\{\Psuk,\Rsetuk\}$, as well as the associated optimum allocation
of $P - \Psuk$ to $\{\Piuk | \Relayi\in\Rsetuk\}$,
constitute the optimum solution to maximizing $\SNRuk$ given $P$.

To facilitate derivation, we assume the relays in $\Rsetuk$ are sorted in the increasing
order of $\Gstik$, and the first relay in $\Rsetuk$ is
$\SRelayki{\buk}$, namely the $b$-th relay in $\SRsetk$.
This means that $\Rsetuk$ is a subset of $\SRsetkitoN{\buk}$,
and $\Gxtouk{\Rsetuk}\leq\Gxtouk{\SRsetkitoN{\buk}}$.
It is important to note that $\Gstoxk{\SRelayki{i}}$ is an increasing function of $i$,
while $\Gxtouk{\SRsetkitoN{i}}$ is a decreasing function of $i$.

Now, $\SNRminRsetuk =\Psuk\Gstoxk{\SRelayki{\buk}}$,
and according to the Schwartz Inequality,
{\small
\begin{align}
   \SNRumrck \leq  \Psuk\Gstuk + (P-\Psuk)\Gxtouk{\Rsetuk}, \label{eq:Schwartz-1}
\end{align}}
where the equality holds when
\begin{align}\label{eq:Schwartz-2}
  \forall\;\Relayi\in\Rsetuk, \;\Piuk =\frac{(P-\Psuk)\Gituk}{\sum_{\Relayi\in\Rsetuk}\Gituk}.
% \forall i,j, (i\neq{j}), \frac{\Piuk}{\Gituk}=\frac{P_{{\rm r}_j,u}(k)}{G_{{\rm r}_j,u}(k)}.
\end{align}

\begin{figure}
  \centering
  \subfigure[case 1]{   %$\buk<\xuk$
     \includegraphics[width=1.3in]{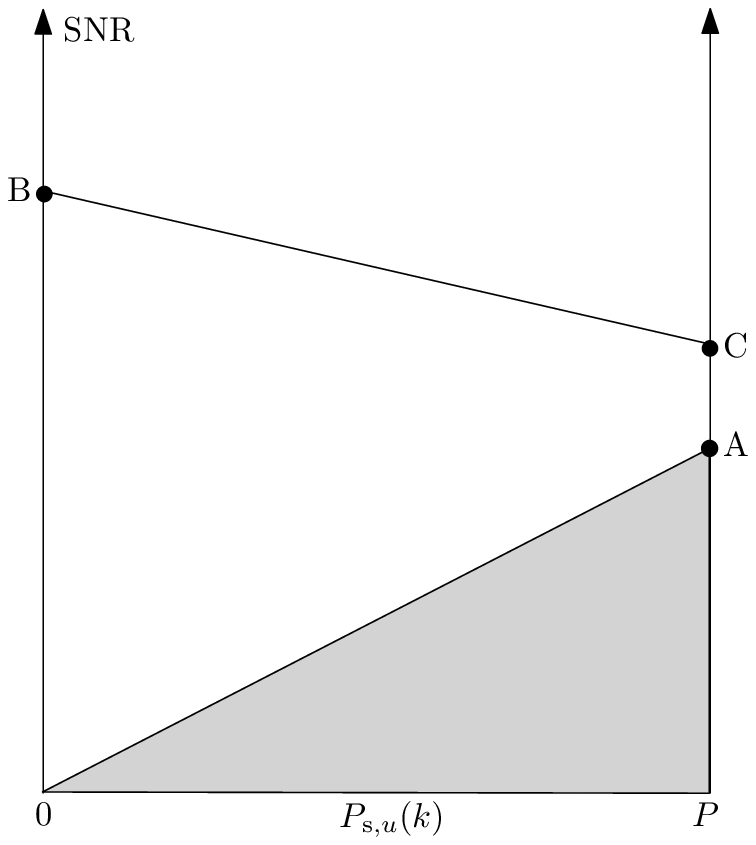}}
  \subfigure[case 2]{    %$\buk>\xuk$, $\Gxtouk{\Rsetuk}\leq\Gstuk$
     \includegraphics[width=1.3in]{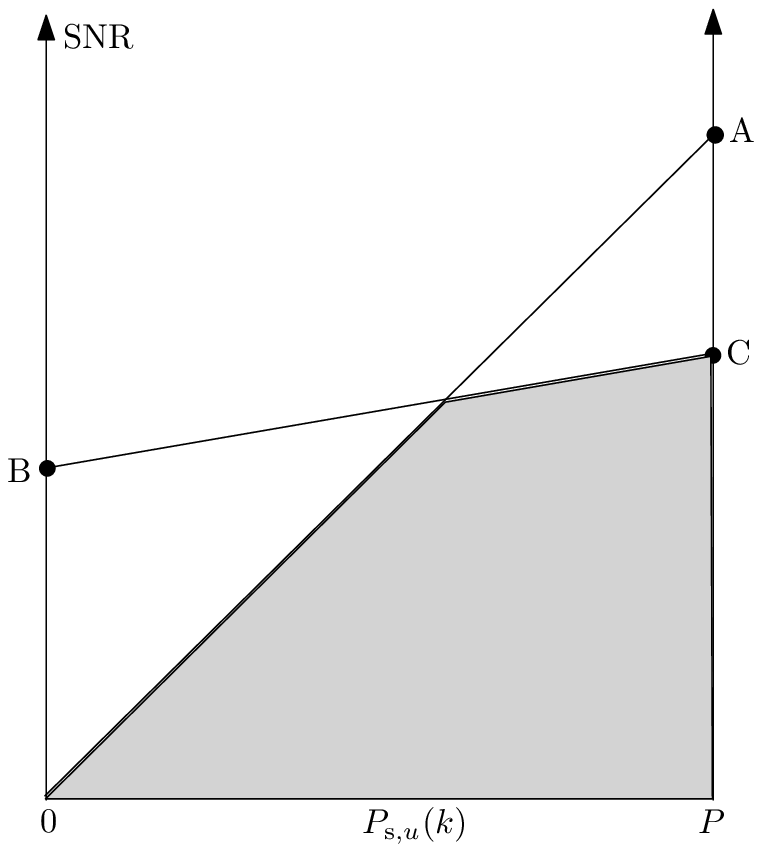}}
  \subfigure[case 3]{%[$\buk>\xuk$, $\Gxtouk{\Rsetuk}>\Gstuk$]
     \includegraphics[width=1.4in]{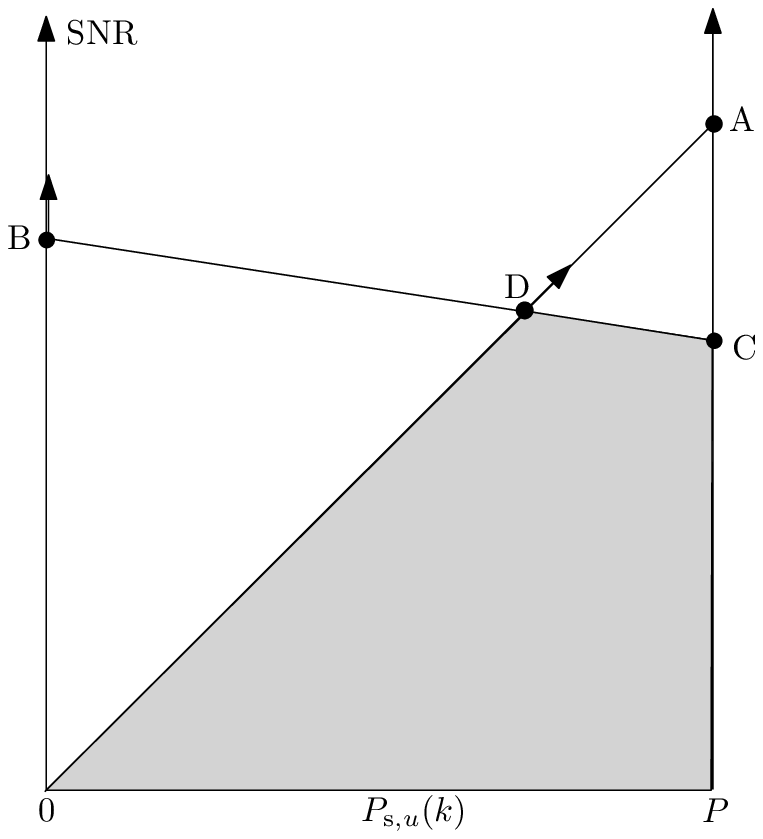}}
   \caption{Illustrations of $\MaxMRCuk$, $\SNRminRsetuk$, and $\MaxMinuk$ versus $\Psuk$.
            $\MaxMinuk$ is highlighted by painting the area it bounds in light grey.}
  \label{fig:Maxmin}
\end{figure}

Obviously, the right hand side of \eqref{eq:Schwartz-1} is equal to $\MaxMRCuk$.
As $\Psuk$ increases from $0$ to $P$,
$(\Psuk,\SNRminRsetuk)$ moves along the line from the origin to A,
whereas $(\Psuk,\MaxMRCuk)$ along the line from B to C
in the coordinate space shown in Figure \ref{fig:Maxmin},
where the coordinates of A, B, and C are given in Table \ref{tab:ABC}.
Note that C is fixed, while both A and B depend on $\Rsetuk$.
In particular, with $\Rsetuk$ and $P$ fixed,
the maximum of $\min\{\MaxMRCuk,\SNRminRsetuk\}$ with respect to $\Psuk$,
denoted by $\MaxMinuk$,
is achieved at A, C, and D for a given $\Rsetuk$ satisfying respectively
\begin{enumerate}
\item
$\buk<\xuk$, where $\xuk$ is the smallest $i$ satisfying $\Gstoxk{\SRelayki{i}}>\Gstuk$.
\item
$\buk\geq\xuk$ and $\Gxtouk{\Rsetuk}\leq\Gstuk$,
\item
$\buk\geq\xuk$ and $\Gxtouk{\Rsetuk}>\Gstuk$.
\end{enumerate}

We can see that only in case $3$, $\MaxMinuk$ is greater than $\Gstuk{P}$.
Among all $\Rsetuk$'s having the same first relay as $\SRelayki{\buk}$ and corresponding to case $3$,
the $\Rsetuk$ with the maximum $\Gxtouk{\Rsetuk}$ corresponds to the greatest height of $D$,
which suggests that $\SRsetkitoN{\buk}$ is the one achieving the maximum $\MaxMinuk$ in that set.
The corresponding $\MaxMinuk$ and $\Psuk$ are computed respectively as
\begin{align}
    \MaxMinuk = \frac{\Gstoxk{\SRelayki{\buk}}\Gxtouk{\SRsetkitoN{\buk}}P}
              {\Gxtouk{\SRsetkitoN{\buk}} + \Gstoxk{\SRelayki{\buk}} - \Gstuk}   \label{Maxminuk}
\end{align}
and
\begin{align}
    \Psuk =  \frac{\Gxtouk{\SRsetkitoN{\buk}}}
              {\Gxtouk{\SRsetkitoN{\buk}} + \Gstoxk{\SRelayki{\buk}} - \Gstuk}P.  \label{eq:PsukR}
\end{align}

Based on the above analysis, we can decide what the optimum $\Rsetuk$, $\Psuk$, and $\Piuk$ should be.
When $\Gxtouk{\SRsetkitoN{\xuk}}\leq\Gstuk$,
every possible $\Rsetuk$ corresponds to either case $1$ or $2$,
which means that the $\MaxMinuk$ maximized over $\Rsetuk$ and $\Psuk$ is equal to $\Gstuk{P}$.
In this case, the maximum $\RukR$ is $\RukR(P) = \ln(1 + \Gstuk{P})$,
and to achieve it $\Psuk$ should be equal to $P$.

When $\Gxtouk{\SRsetkitoN{\xuk}}>\Gstuk$,
every possible $\Rsetuk$ corresponds to one of cases $1$, $2$, and $3$.
Obviously, the optimum $\Rsetuk$ should correspond to case $3$.
To have $\Rsetuk$ corresponding to case $3$, $\buk$ must be between $\xuk$ and $\yuk$,
where $\yuk$ is the greatest $i$ satisfying $\Gxtouk{\SRsetkitoN{i}}>\Gstuk$.
As pointed out before,
$\SRsetkitoN{\buk}$ is the $\Rsetuk$ achieving the maximum $\MaxMinuk$
among all $\Rsetuk$'s having the same first relay as $\SRelayki{\buk}$.
Therefore, the optimum $\Rsetuk$ should be $\SRsetkitoN{\zuk}$,
where $\zuk$ is the $\buk$ between $\xuk$ and $\yuk$ and maximizing $\MaxMinuk$,
namely the maximizer of the right-hand side of \eqref{eq:Gu1k}.
In this case, the maximum $\RukR$ and the optimum RA can be evaluated with the formulas
for the third case shown in Section III.B.

\begin{table}
  \centering
  \caption{Coordinates of A, B, and C.}\label{tab:ABC}
  \begin{tabular}{|c|c|c|c|}
    \hline
               &    A                              &  B                        &  C            \\
    \hline
     $\Psuk$   &  $P$                              & $0$                       &  $P$          \\
    \hline
     SNR       & $\Gstoxk{\SRelayki{\buk}}{P}$     & $\Gxtouk{\Rsetuk}{P}$     & $\Gstuk{P}$   \\
    \hline
  \end{tabular}
\end{table}

\section*{Acknowledgement}

The authors would like to thank Prof. Josep Vidal for coordinating the review,
as well as the anonymous reviewers for their valuable comments and suggestions
to improve the quality of this paper.

\ifCLASSOPTIONcaptionsoff
  \newpage
\fi



{\small
\bibliographystyle{IEEEtran}
\bibliography{Relay-sum-bib}

\begin{thebibliography}{10}
\providecommand{\url}[1]{#1}
\csname url@samestyle\endcsname
\providecommand{\newblock}{\relax}
\providecommand{\bibinfo}[2]{#2}
\providecommand{\BIBentrySTDinterwordspacing}{\spaceskip=0pt\relax}
\providecommand{\BIBentryALTinterwordstretchfactor}{4}
\providecommand{\BIBentryALTinterwordspacing}{\spaceskip=\fontdimen2\font plus
\BIBentryALTinterwordstretchfactor\fontdimen3\font minus
  \fontdimen4\font\relax}
\providecommand{\BIBforeignlanguage}[2]{{%
\expandafter\ifx\csname l@#1\endcsname\relax
\typeout{** WARNING: IEEEtran.bst: No hyphenation pattern has been}%
\typeout{** loaded for the language `#1'. Using the pattern for}%
\typeout{** the default language instead.}%
\else
\language=\csname l@#1\endcsname
\fi
#2}}
\providecommand{\BIBdecl}{\relax}
\BIBdecl

\bibitem{Pabst04}
R.~Pabst, B.~Walke, D.~Schultz \emph{et~al.}, ``Relay-based deployment concepts
  for wireless and mobile broadband radio,'' \emph{IEEE Communications
  Magazine}, vol.~42, no.~9, pp. 80--89, sept. 2004.

\bibitem{Laneman04}
J.~N. Laneman, D.~N.~C. Tse, and G.~W. Wornell, ``Cooperative diversity in
  wireless networks: efficient protocols and outage behavior,'' \emph{IEEE
  Transactions on Information Theory}, vol.~50, no.~12, pp. 3062--3080, Dec.
  2004.

\bibitem{Wang11}
T.~Wang and L.~Vandendorpe, ``Iterative resource allocation for maximizing
  weighted sum min-rate in downlink cellular {OFDMA} systems,'' \emph{IEEE
  Transactions on Signal Processing}, vol.~59, no.~1, pp. 223--234, Jan. 2011.

\bibitem{Gui08}
B.~Gui and L.~J. Climini-Jr, ``Bit loading algorithms for cooperative {OFDM}
  systems,'' \emph{Eurasip Journal on Wireless Communications and Networking},
  2008.

\bibitem{Li08}
Y.~Li, W.~Wang, J.~Kong \emph{et~al.}, ``Power allocation and subcarrier
  pairing in {OFDM}-based relaying networks,'' in \emph{IEEE International
  Conference on Communications}, 2008, pp. 2602--2606.

\bibitem{Vandendorpe08-1}
L.~Vandendorpe, R.~Duran, J.~Louveaux \emph{et~al.}, ``Power allocation for
  {OFDM} transmission with {DF} relaying,'' in \emph{IEEE International
  Conference on Communications}, 2008, pp. 3795 --3800.

\bibitem{Vandendorpe08-2}
L.~Vandendorpe, J.~Louveaux, O.~Oguz \emph{et~al.}, ``Improved {OFDM}
  transmission with {DF} relaying and power allocation for a sum power
  constraint,'' in \emph{ISWPC}, 2008, pp. 665--669.

\bibitem{Vandendorpe09-1}
------, ``Power allocation for improved {DF} relayed {OFDM} transmission: the
  individual power constraint case,'' in \emph{IEEE International Conference on
  Communications}, 2009, pp. 1--6.

\bibitem{Vandendorpe09-2}
------, ``Rate-optimized power allocation for {DF}-relayed {OFDM} transmission
  under sum and individual power constraints,'' \emph{Eurasip Journal on
  Wireless Communications and Networking}, vol. 2009.

\bibitem{Vandendorpe09-3}
------, ``Rate-optimized power allocation for {OFDM} transmission with multiple
  {DF}/regenerative relays and an improved protocol,'' in \emph{IEEE Wireless
  Communications and Networking Conference}, 2009, pp. 1--6.

\bibitem{Vandendorpe09-4}
------, ``Rate-optimized power allocation for {OFDM} transmission with multiple
  {DF} relays and individual power constraints,'' in \emph{European Signal
  Processing Conference}, 2009, pp. 145--149.

\bibitem{WangJSAC11}
T.~Wang and L.~Vandendorpe, ``Sum rate maximized resource allocation in
  multiple {DF} relays aided {OFDM} transmission,,'' \emph{to appear in IEEE
  Journal on Selected Areas in Communications}, Sept. 2011.

\bibitem{Nam07}
W.~Nam, W.~Chang, S.-Y. Chung \emph{et~al.}, ``Transmit optimization for
  relay-based cellular {OFDMA} systems,'' in \emph{IEEE International
  Conference on Communications}, 2007, pp. 5714 --5719.

\bibitem{Kaneko07}
M.~Kaneko and P.~Popovski, ``Radio resource allocation algorithm for
  relay-aided cellular {OFDMA} system,'' in \emph{IEEE International Conference
  on Communications}, 2007, pp. 4831 --4836.

\bibitem{Kwak07}
R.~Kwak and J.~Cioffi, ``Resource-allocation for {OFDMA} multi-hop relaying
  downlink systems,'' in \emph{IEEE Global Telecommunications Conference},
  2007, pp. 3225 --3229.

\bibitem{Cui09}
Y.~Cui, V.~K.~N. Lau, and R.~Wang, ``Distributive subband allocation, power and
  rate control for relay-assisted {OFDMA} cellular system with imperfect system
  state knowledge,'' \emph{IEEE Transactions on Wireless Communications},
  vol.~8, no.~10, pp. 5096--5102, Oct. 2009.

\bibitem{Salem10}
M.~Salem, A.~Adinoyi, M.~Rahman \emph{et~al.}, ``Fairness-aware radio resource
  management in downlink {OFDMA} cellular relay networks,'' \emph{IEEE
  Transactions on Wireless Communications}, vol.~9, no.~5, pp. 1628 --1639, May
  2010.

\bibitem{Ng07}
T.~C.-Y. Ng and W.~Yu, ``Joint optimization of relay strategies and resource
  allocations in cooperative cellular networks,'' \emph{IEEE Journal on
  Selected Areas in Communications}, vol.~25, no.~2, pp. 328 --339, Feb. 2007.

\bibitem{Kadloor10}
S.~Kadloor and R.~Adve, ``Relay selection and power allocation in cooperative
  cellular networks,'' \emph{IEEE Transactions on Wireless Communications},
  vol.~9, no.~5, pp. 1676--1685, May 2010.

\bibitem{Vardhe10}
K.~Vardhe, D.~Reynolds, and B.~D. Woerner, ``Joint power alllocation and relay
  selection for multiuer cooperative communications,'' \emph{IEEE Transactions
  on Wireless Communications}, vol.~9, no.~4, pp. 1255--1260, Apr. 2010.

\bibitem{OFDMAsyn07}
M.~Morelli, C.~Kuo, and M.~Pun, ``Synchronization techniques for orthogonal
  frequency division multiple access (ofdma): a tutorial review,''
  \emph{Proceeding of IEEE}, vol.~95, no.~7, pp. 1394--1427, July 2007.

\bibitem{Fund-WCOM}
D.~Tse and P.~Viswanath, \emph{Fundamentals of Wireless Communication}.\hskip
  1em plus 0.5em minus 0.4em\relax Cambridge University Press, 2005.

\bibitem{Convex-opt}
S.~Boyd and L.~Vandenberghe, \emph{Convex optimization}.\hskip 1em plus 0.5em
  minus 0.4em\relax Cambridge University Press, 2004.

\bibitem{Nonlinear-opt}
D.~P. Bertsekas, \emph{Nonlinear programming, $2$nd edition}.\hskip 1em plus
  0.5em minus 0.4em\relax Athena Scientific, 2003.

\bibitem{Yu06}
W.~Yu and R.~Lui, ``Dual methods for nonconvex spectrum optimization of
  multicarrier systems,'' \emph{IEEE Transactions on Communications}, vol.~54,
  no.~7, pp. 1310--1322, July 2006.

\end{thebibliography}
}

%\begin{thebibliography}{1}


%\end{thebibliography}

\begin{IEEEbiography}[{\includegraphics[width=1in,height=1.5in,clip,keepaspectratio]{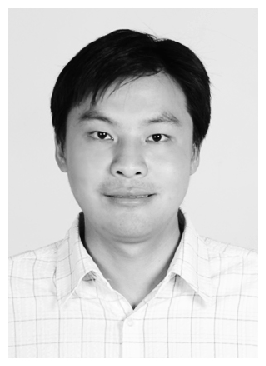}}]{Tao Wang}
received respectively B.E. ({\it summa cum laude}) and DoE degree in electronic engineering
from Zhejiang University, China, in 2001 and 2006,
as well as civil electrical engineering degree ({\it summa cum laude})
from Universit{\'e} Catholique de Louvain (UCL), Belgium, in 2008.
He had multiple research appointments
in Motorola Electronics Ltd. Suzhou Branch, China, since 2000 to 2001,
the Institute for Infocomm Research (I$^2$R), Singapore, since 2004 to 2005,
and Delft University of Technology and Holst Center in the Netherlands since 2008 to 2009.
He is with UCL since 2009.
He has been an associate editor-in-chief for {\it Signal Processing: An International Journal (SPIJ)} since Oct. 2010.
His current research interests are in the optimization of wireless localization systems
with energy awareness, as well as resource allocation algorithms in wireless communication systems.
\end{IEEEbiography}

\begin{IEEEbiography}[{\includegraphics[width=1in,height=1.5in,clip,keepaspectratio]{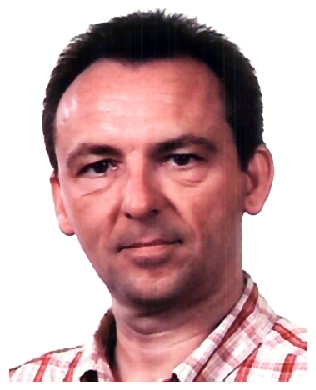}}]{Luc Vandendorpe}
was born in Mouscron, Belgium in 1962. He received the Electrical Engineering degree ({\it summa cum laude})
and the Ph. D. degree from the Universit{\'e} Catholique de Louvain (UCL) Louvain-la-Neuve, Belgium
in 1985 and 1991 respectively. Since 1985, he is with the Communications and Remote
Sensing Laboratory of UCL where he first worked in the field of bit rate reduction techniques
for video coding. In 1992, he was a Visiting Scientist and Research Fellow at the Telecommunications
and Traffic Control Systems Group of the Delft Technical University, The Netherlands, where he worked
on Spread Spectrum Techniques for Personal Communications Systems.
From October 1992 to August 1997, L. Vandendorpe was Senior Research Associate of the Belgian
NSF at UCL, and invited assistant professor. Presently he is Professor and head of
the Institute for Information and Communication Technologies, Electronics and Applied Mathematics.

His current interest is in digital communication systems and more precisely resource
allocation for OFDM(A) based multicell systems, MIMO and distributed MIMO,
sensor networks, turbo-based communications systems, physical layer security and UWB based positioning.
In 1990, he was co-recipient of the Biennal Alcatel-Bell Award from the Belgian NSF
for a contribution in the field of image coding. In 2000 he was co-recipient
(with J. Louveaux and F. Deryck) of the Biennal Siemens Award from the Belgian NSF
for a contribution about filter bank based multicarrier transmission.
In 2004 he was co-winner (with J. Czyz) of the Face Authentication Competition,
FAC 2004. L. Vandendorpe is or has been TPC member for numerous IEEE conferences
(VTC Fall, Globecom Communications Theory Symposium, SPAWC, ICC) and for the Turbo Symposium.
He was co-technical chair (with P. Duhamel) for IEEE ICASSP 2006.
He was an editor of the IEEE Trans. on Communications for Synchronisation and
Equalization between 2000 and 2002, associate editor of the IEEE Trans. on
Wireless Communications between 2003 and 2005, and associate editor of the
IEEE Trans. on Signal Processing between 2004 and 2006. He was chair of the
IEEE Benelux joint chapter on Communications and Vehicular Technology
between 1999 and 2003. He was an elected member of the Signal Processing for
Communications committee between 2000 and 2005, and an elected member of the Sensor
Array and Multichannel Signal Processing committee of the Signal Processing Society between 2006 and 2008.
Currently, he is an elected member of the Signal Processing for Communications committee.
He is the Editor in Chief for the Eurasip Journal on Wireless Communications and Networking.
He is a Fellow of the IEEE.
\end{IEEEbiography}

%\vfill


\end{document}